\documentclass[fleqn,usenatbib]{mnras}

\usepackage[T1]{fontenc}
\usepackage{ae,aecompl}

\DeclareRobustCommand{\VAN}[3]{#2}
\let\VANthebibliography\thebibliography
\def\thebibliography{\DeclareRobustCommand{\VAN}[3]{##3}\VANthebibliography}

\usepackage{graphicx}	
\usepackage{amsmath}	
\usepackage{amssymb}	
\usepackage{color}
\usepackage{algorithm}
\usepackage{booktabs}
\usepackage[noend]{algpseudocode}
\usepackage{pifont}
\usepackage{soul}
\usepackage{float}
\usepackage{multibib}
\newcites{Appendix}{Appendix References}
\usepackage{natbib}
\usepackage[flushleft]{threeparttablex}
\usepackage{mathrsfs}
\usepackage{widetext}
\usepackage{mathtools}
\usepackage{tikz}
\usepackage{esint}
\usepackage[draft]{todonotes}
\usepackage{lipsum}
\usepackage{longtable}
\usepackage{pifont}

\usetikzlibrary{calc,shapes}

\graphicspath{{Figures/}}

\newcommand{\appropto}{\mathrel{\vcenter{
  \offinterlineskip\halign{\hfil$##$\cr
    \propto\cr\noalign{\kern2pt}\sim\cr\noalign{\kern-2pt}}}}}

\newcommand{\vecB}[1]{\mathrm{{\bmath{\mathit{#1}}}}}
\newcommand{\Exp}[2]{\left\langle{#1}\right\rangle_{#2}}

\renewcommand{\d}[1]{\ensuremath{\operatorname{d}\!{#1}}}
\newcommand{\dthree}[1]{\ensuremath{\operatorname{d}^3\!{#1}}}

\DeclareMathOperator\V{\mathcal{V}}
\newcommand{\M}{\mathcal{M}}
\DeclareMathOperator\Ma{\mathcal{M}_{\text{A}}}

\newcommand\ekin{E_{\rm kin}}
\newcommand\ekinb{\Exp{\ekin}{\V}}
\newcommand\emag{E_{\rm mag}}
\newcommand\emagb{\Exp{\emag}{\V}}

\DeclareMathAlphabet\mathbfcal{OMS}{cmsy}{b}{n}

\newcommand{\Rm}{\text{Rm}}
\renewcommand{\Re}{\text{Re}}
\newcommand{\Pm}{\text{Pm}}
\renewcommand{\v}{\mathit{v}}
\renewcommand{\b}{\mathit{b}}
\newcommand{\kcor}{k_{\text{cor}}}
\newcommand{\keta}{k_\eta}
\newcommand{\keq}{k_{\text{eq}}}

\newcommand{\ku}{k_{\text{u}}}
\newcommand{\kp}{k_{\text{peak}}}

\newcommand{\lo}{\ell_0}
\newcommand{\lnu}{\ell_\nu}
\newcommand{\leta}{\ell_\eta}

\defcitealias{Brunt2010a}{BFP2010}
\defcitealias{Beattie2020}{BF2020}
\defcitealias{Kolmogorov1941}{K41}
\defcitealias{Menon2020}{MFK2020}
\defcitealias{Hopkins2013}{H2013}
\defcitealias{Skalidis2021}{S+2021}

\usepackage{scalerel,tikz}
\usetikzlibrary{svg.path}
\definecolor{orcidlogocol}{HTML}{A6CE39}
\tikzset{orcidlogo/.pic={\fill[orcidlogocol] svg{M256,128c0,70.7-57.3,128-128,128C57.3,256,0,198.7,0,128C0,57.3,57.3,0,128,0C198.7,0,256,57.3,256,128z}; \fill[white] svg{M86.3,186.2H70.9V79.1h15.4v48.4V186.2z} svg{M108.9,79.1h41.6c39.6,0,57,28.3,57,53.6c0,27.5-21.5,53.6-56.8,53.6h-41.8V79.1z M124.3,172.4h24.5c34.9,0,42.9-26.5,42.9-39.7c0-21.5-13.7-39.7-43.7-39.7h-23.7V172.4z} svg{M88.7,56.8c0,5.5-4.5,10.1-10.1,10.1c-5.6,0-10.1-4.6-10.1-10.1c0-5.6,4.5-10.1,10.1-10.1C84.2,46.7,88.7,51.3,88.7,56.8z};}}
\newcommand\orcidicon[1]{\href{https://orcid.org/#1}{\mbox{\scalerel*{
\begin{tikzpicture}[yscale=-1,transform shape]\pic{orcidlogo};
\end{tikzpicture}}{|}}}}

\newcommand{\aref}[1]{\hyperref[#1]{Appendix~\ref{#1}}}
\newcommand{\viz}{\textit{viz.}}
\newcommand{\nquad}[1][1]{\hspace*{#1em}\ignorespaces}

\title[G or D -- I: saturation]{Growth or Decay -- I: universality of the turbulent dynamo saturation}

\author[Beattie, et al., 2023]{
James R. Beattie$^{\orcidicon{0000-0001-9199-7771}\,1,2}$\thanks{E-mail: james.beattie@anu.edu.au}, 
Christoph Federrath$^{\orcidicon{0000-0002-0706-2306}\,1,3}$,
Neco Kriel$^{\orcidicon{0000-0002-3558-3926}\,1}$,
Philip Mocz$^{\orcidicon{0000-0001-6631-2566}\,4,5}$,
and Amit Seta$^{\orcidicon{0000-0001-9708-0286}\,1}$
\\
$^{1}$Research School of Astronomy and Astrophysics, Australian National University, Canberra, ACT 2611, Australia \\
$^{2}$Department of Astronomy and Astrophysics, University of California, Santa Cruz, 1156 High Street, Santa Cruz, CA 96054\\
$^{3}$Australian Research Council Centre of Excellence in All Sky Astrophysics (ASTRO3D), Canberra, ACT 2611, Australia \\
$^{4}$Department of Astrophysical Sciences, Princeton University, 4 Ivy Lane, Princeton, NJ 08544, USA \\
$^{5}$Lawrence Livermore National Laboratory, 7000 East Ave, Livermore, CA, USA \\
}

\date{Accepted XXX. Received YYY; in original form ZZZ}

\pubyear{2023}

\begin{document}
\label{firstpage}
\pagerange{\pageref{firstpage}--\pageref{lastpage}}
\maketitle

\begin{abstract}
    The turbulent small-scale dynamo (SSD) is likely to be responsible for the magnetisation of the interstellar medium (ISM) that we observe in the Universe today. The SSD efficiently converts kinetic energy $\ekin$ into magnetic energy $\emag$, and is often used to explain how an initially weak magnetic field with $\emag \ll \ekin$ is amplified, and then maintained at a level $\emag \lesssim \ekin$. Usually, this process is studied by initialising a weak seed magnetic field and letting the turbulence grow it to saturation. However, in this Part I of the Growth or Decay series, using three-dimensional, visco-resistive magnetohydrodynamical turbulence simulations up to magnetic Reynolds numbers of 2000, we show that the same final state in the integral quantities, energy spectra, and characteristic scales of the magnetic field can also be achieved if initially $\emag \sim \ekin$ or even if initially $\emag \gg \ekin$. This suggests that the final saturated state of the turbulent dynamo is set by the turbulence and the material properties of the plasma, independent of the initial structure or amplitude of the magnetic field. We discuss the implications this has for the maintenance of magnetic fields in turbulent plasmas and future studies exploring the dynamo saturation.
\end{abstract}


\section{Introduction}\label{sec:intro}
    \subsection{The turbulent dynamo}   
    
    The present day Universe is magnetised, thus beckoning the question: how did it become so? One answer is the turbulent small-scale dynamo (SSD; or fluctuation dynamo), which is a mechanism for transforming turbulent kinetic energy into turbulent magnetic energy until both are statistically stationary and approximately in energy equipartition -- the so-called saturated phase of the SSD. In the interstellar medium (ISM) of galaxies, saturation probably occurred at redshifts $z \approx 25-8$, depending upon the nature of density fluctuations \citep{Xu2016_dynamo,McKee2020}, consequentially making the study of present-day ISM magnetic fields the study of the saturated stage of the SSD. However, understanding both the physics, statistics, and constructing a predictive model for the saturation of the SSD remains an active problem in the dynamo community. 
    
    In a Markovian fashion, the saturated state of the magnetic field that develops from the SSD forgets the field that seeded it \citep{Seta2020_seed_magnetic_field}. These initial, primordial fields may have been incredibly weak (maybe as weak as $10^{-16}\,\rm{G}$ in the gas of the intergalactic medium), and perhaps were generated from a battery process \citep[e.g., ][]{Biermann1950_battery} or through spontaneous magnetic field creation during the electroweak symmetry breaking epoch in the primordial Universe \citep[e.g., ][]{Brandenburg2005_dynamo_review,Subramanian2016_origins_review,Subramanian2019_origins}. Such a weak, fluctuating seed field $\vecB{b}$ can be amplified exponentially fast in time, \viz, $\Exp{\b^2}{\V} \propto \exp\left\{ \gamma t \right\}$ by turbulent motions in the plasma, where $\Exp{\b^2}{\V}$ is the system volume ($\V\equiv L^3$) integral magnetic energy and $\gamma \sim v_{\nu}/\lnu$ is the growth rate that goes with the reciprocal dynamical time of the turbulence at the viscous scale $\lnu$ \citep[e.g.,][]{McKee2020}. This is termed the kinematic, exponential growth, or linear induction equation stage of the turbulent dynamo \citep[e.g.,][]{Brandenburg2005_dynamo_review}, and lasts until $\Exp{\b^2}{\V}$ becomes strong enough to cause a backreaction on the momentum transport through $(\nabla\times\vecB{b})\times\vecB{b} \equiv \nabla\cdot(\vecB{b}\otimes\vecB{b}) - (1/2)\nabla b^2$,\footnote{where $\otimes$ is the tensor product, e.g., $\nabla\cdot(\vecB{b} \otimes \vecB{b}) = \partial_ib_ib_j$, in Einstein notation.} e.g., when $\nabla\cdot(\vecB{v}\otimes\vecB{v}) \sim \nabla\cdot(\vecB{b}\otimes\vecB{b})$ \citep{Schekochihin2004_dynamo,Galishnikova2022_saturation_and_tearing}. In this stage, the induction equation becomes strongly nonlinear because $\vecB{v}$ satisfies the momentum equation with non-negligible $(\nabla\times\vecB{b})\times\vecB{b}$, which then contributes to the induction equation and the overall time-evolution of $\vecB{b}$. This stage is termed the linear growth, because $\Exp{\b^2}{\V} \propto t$, or the nonlinear induction equation stage of the dynamo \citep[e.g.,][]{Schekochihin2002_saturation_evolution,Maron2004_nonlinear_magnetic_spectrum,Cho2009_linear_growth,Xu2016_dynamo}. Finally, after a sufficient amount of time $\Exp{\b^2}{\V}$ saturates, such that\footnote{Note that previous works suggested $\Exp{\b^2}{\V} \sim \Re^{-1/2}\Exp{\v^2}{\V}$ \citep{Batchelor1950_dynamos}. This would result in an exceptionally weak magnetic field for astrophysical media, such as molecular clouds in the interstellar medium, which are characterised by $\Re$ as high as $\Re \sim 10^9$ \citep{Krumholz2014}. However SSD simulations and now laboratory experiments \citep{Tzeferacos2018_dynamo_in_the_lab,Liao2019_dynamo_in_the_lab} show higher levels of saturation \citep[see \S3.2.1 in][for more details]{Rincon2019_dynamo_theories}.} $\Exp{\b^2}{\V} \sim \Exp{\v^2}{\V}$. The integral quantities of the saturation depend upon the compressibility of the plasma (e.g., the turbulent Mach number $\M = \Exp{v^2}{\V}^{1/2}/c_s$, where $c_s$ is the sound speed; see \citealt{Haugen2004_dynamo_mach_and_crit_rm,Federrath2011_mach_dynamo,Seta2020_saturation_high_Pm,seta2021saturation}), the nature of the turbulent driving source \citep{Federrath2011_mach_dynamo,Chirakkara2021}, the diffusion timescales for the magnetic and velocity fields -- the magnetic Prandtl number, $\Pm$ (for finite $\Pm$) \citep[e.g.,][]{Schober2012_saturation_Re_Rm_dependence}, and the thermodynamic phase structure of the plasma \citep{Seta2022_multiphase_dynamo,Gent2022_multiphase_dynamo}.
    
    Numerical experiments have shown that $\Exp{\b^2}{\ell^3} \sim \Exp{\v^2}{\ell^3}$ need not be true for all $\ell$ in the magnetised plasma, and the saturation of the SSD is a scale-dependent (saturation looks different on different Fourier modes) phenomenon \citep[e.g.,][]{Schekochihin2002_saturation_evolution,Maron2004_nonlinear_magnetic_spectrum,Schober2012_saturation_Re_Rm_dependence,Schober2015_saturation_of_turbulent_dynamo}. Here we summarise the phenomenology championed most recently by \citet{Galishnikova2022_saturation_and_tearing}. As the dynamo approaches saturation, the magnetic field fluctuations on each scale $b_{\ell}^2$ reach energy equipartition with the turbulence on that scale $v_{\ell}^2 \sim b_{\ell}^2$, successively moving the energy equipartition scale (equivalent to the shearing scale in \citealt{Maron2004_nonlinear_magnetic_spectrum})  $\ell_{\rm eq} \sim k_{\rm eq}^{-1}$ from small, viscous-dominated scales $\ell_{\rm eq} \sim \lnu$ (where $t_{\ell} = \ell/ v_{\ell}$ is short), to the larger and slower eddies on $\ell \gg \lnu$ until a maximal stretching rate is achieved, given by $t_{\text{max}}^{-1} \sim \big(\Exp{\v^2}{\V}^{1/2}/L\big)\big(\Exp{\v^2}{\V}^{1/2}/\Exp{\b^2}{\V}^{1/2}\big)^2$ on $\ell_{\rm eq}$ \citep{Galishnikova2022_saturation_and_tearing}. At this point, only $\ell > \ell_{\rm eq}$ are not suppressed by the magnetic tension, and hence the final value of $\Exp{\b^2}{\V}/\Exp{\v^2}{\V}$ is sensitive to where this scale is. Clearly, for supersonic dynamo experiments, where $\Exp{\b^2}{\V}/\Exp{\v^2}{\V}$ is reduced compared to the simulated subsonic counterparts \citep[see for example,][]{Federrath2011_mach_dynamo,seta2021saturation} and $\ell_{\rm eq}$ must be on smaller scales, which allows for hydrodynamic motions to dominate over a larger range of $k$. The details of the exact saturation mechanism is still, however, an active area of research \citep[some ideas and discussion in][]{Rincon2019_dynamo_theories, Seta2020_saturation_high_Pm,seta2021saturation}. 
    
    \subsection{Strong magnetic fields decaying into driven turbulence}

    In classical dynamo experiments, initial $\vecB{b}$ fields are set such that $\Exp{\b^2}{\V}/\Exp{\v^2}{\V} \ll 1$. The turbulence is driven, and through the conversion of $\v^2$ into $\b^2$, $\vecB{b}$ grows. However, what happens if $\Exp{\b^2}{\V}/\Exp{\v^2}{\V} \gg 1$ to begin with, i.e., the initial magnetic energy is in superequipartition with the kinetic energy? In the ISM, this may be realised through large-scale compressions of the plasma through galaxy-galaxy interactions or radial flows boosting the magnetic field \citep{Steinwandel2020_galactic_winds_beyond_equipartition_B}, and small-scale events such as supernova driven shocks \citep[e.g.,][]{Korpi1999_supernova_regulated_ISM,Lu2020_supernova_driving,Chevance2022_supernova_time_delay}, or stellar feedback \citep[e.g.,][]{Lancaster2021_stellar_winds,Menon2022_stellar_IR_feedback}. Such compressions may enhance a magnetic field on scales where flux-freezing holds (valid above the scales of ion-neutral damping, $\mathcal{O}(10^{-3}-10^{-2}\,\rm{pc})$ for Alfv\'en modes; \citealt{Krumholz2020}), such that $B \propto \rho^\alpha$, where $\rho$ is the gas density and $\alpha$ is the enhancement factor that depends upon the detailed geometry of the compression \citep{Tritsis2016,Mocz2018,Beattie2021}. In these circumstances, the superequipartition magnetic field must decay into a lower energy state, which may be (but not necessarily) the same saturated state as set by the small-scale dynamo, i.e., it is not clear if different, strong magnetic fields (in our case, with a different magnetic morphology) intrinsically change the $t \rightarrow \infty$ behaviour of the turbulent magnetic field. Furthermore, it is not clear if the properties of the saturated magnetic field are functions of the detailed workings in the kinematic and non-linear stages of the SSD or indifferent to them. For example, is the folded magnetic field structure developed in the kinematic stage critical for the saturation statistics, as suggested in \citet{Galishnikova2022_saturation_and_tearing}? Hence, understanding the $t\rightarrow\infty$ state of the integral and spectral properties for a magnetic field initially in superequipartition, bypassing the regular kinematic and nonlinear stages, is the key motivation for this first ``Growth or Decay" study.

    Even though the superequipartition experiments in this series are not typical decaying turbulence calculations, which would generally describe the process where both $\ekin$ and $\emag$ decay simultaneously from an initial condition, we provide a short discussion of decaying turbulence theory, which we expand upon in Paper II alongside a detailed study of the decay process itself, including the length and timescales involved in decay, and the physical processes that determine the superequipartition decay into saturation.
    
    Decaying MHD turbulence is a well-studied process \citep[see ][\S12 for a recent review]{Schekochihin2020_bias_review}, which is relevant to many astrophysical phenomena where the turbulence driving mechanism may be intermittent in space and time. Critical to the MHD decaying phenomenology is the volume-averaged magnetic helicity, $h = \Exp{ \vecB{a}\cdot\vecB{b}}{\V} \sim b^2\ell_0$, where $\vecB{a}$ is the vector potential of $\vecB{b}$, $\vecB{b} = \nabla \times \vecB{a}$. $h$ is a topological invariant of the magnetic field, perfectly conserved as $\eta \rightarrow 0$ and approximately conserved for small (non-vanishing) $\eta$ \citep{Hosking2021_reconnection_controlled_decay}. Hence, in a similar fashion as the Loitsyansky integral \citep{Kolmogorov1941_decaying}, one can construct a number of decay models based on $h$ (and other invariants of MHD plasmas, like cross-helicity, etc.). For example, \citet{Hosking2021_reconnection_controlled_decay} considered the decay of non-helical $\Exp{\b^2}{\V}/\Exp{\v^2}{\V} \gg 1$ turbulence via the Hosking invariant, $I_H = \int\dthree{\vecB{r}}\,\Exp{h(\vecB{x})h(\vecB{x}+\vecB{r})}{\V} \sim b^4 \ell_0^5$. By assuming that the decay is controlled by reconnection rates, they showed that $\emag \sim \ekin \propto t^{-10/9}$ for fast, plasmoid-dominated reconnection \citep[e.g.,][]{Bhattacharjee2009_fast_reconnection,Uzdensky2010_plasmoid_reconnection,Loureiro_2016_plasmoid_instability}, and $\emag \propto t^{-20/17}$ and $\ekin \propto t^{-19/17}$ for slow reconnection \citep{Sweet1958_reconnection, Parker1957_reconnection}. \citet{Zhou2022_testing_Hosking} provided high-resolution, numerical support for these models, highlighting the important role of the Hosking invariant in decaying non-helical MHD turbulence. However, these models assume that $\ekin$ is purely being sourced by reconnection outflows (usually the velocity field is exactly zero in the initial condition for these experiments), which is not the case in our experiments, where $\ekin$ is stochastically driven by large-scale momentum modes in the plasma, which may be suppressed for $\emagb/\ekinb \gg 1$, but for our experiments at $\emagb/\ekinb \lesssim 10^2$ the forcing always plays a role. Moreover, in our parameter regime, we will show that a very different decay process is present, involving the growth and coalescence of sub-Alfv\'enic, helical ropes of magnetic field flux tubes. The flux ropes become force-free ($\vecB{j}\times\vecB{b}=0$), which in turn linearises the induction equation and results in exponential decay, which we study, model in detail, and compare with the models mentioned in this section in Paper II. 

    \begin{table*}
    \caption{Main simulation parameters and derived quantities.}
    \label{tab:sims}
    \begin{center}
    \begin{tabular}{l c c c c c c c c c c c c}
        \hline\hline
                Sim. ID & $\Re$ & $\Rm$ & $\nu t_0/\ell_0^2$ & $\eta t_0/\ell_0^2$ & $\M$ & $b_{\rm init}$ & $\frac{E_{\rm mag,0}}{\ekin}$ & $\left( \frac{\emag}{\ekin} \right)_{\text{sat}}$ & $(\Ma)_{\text{sat}}$ & $N_{\rm grid}^3$  \\[0.4em]
                \multicolumn{1}{c}{(1)} & (2) & (3) & (4) & (5) & (6) & (7) & (8) & (9) & (10) & (11) \\
                \hline
                \multicolumn{11}{c}{$\Pm = 1$}\\
                \hline
                \texttt{weakPm1} & 500 & 500   & $2\times 10^{-3}$ & $2\times 10^{-3}$ & 0.5 & $1 \leq |\vecB{k}L/2\pi|\leq 3$ & $10^{-10}$  & 0.09 $\pm$ 0.03 & 3.43 $\pm$ 0.63 & $288^3$ \\
                \texttt{satPm1} & 500 & 500    & $2\times 10^{-3}$ & $2\times 10^{-3}$ & 0.5 & $1 \leq |\vecB{k}L/2\pi|\leq 3$ & $10^{-2}$   & 0.10 $\pm$ 0.04 & 3.31 $\pm$ 0.62 & $288^3$ \\
                \texttt{strongPm1} & 500 & 500 & $2\times 10^{-3}$ & $2\times 10^{-3}$ & 0.5 & $1 \leq |\vecB{k}L/2\pi|\leq 3$ & $10^{2}$    & 0.10 $\pm$ 0.03 & 3.32 $\pm$ 0.64 & $288^3$ \\
                \texttt{initbPm1} & 500 & 500  & $2\times 10^{-3}$ & $2\times 10^{-3}$ & 0.5 & $7 \leq |\vecB{k}L/2\pi|\leq 9$ & $10^{2}$    & 0.11 $\pm$ 0.04 & 3.13 $\pm$ 0.54 & $288^3$ \\ 
                \hline
                \multicolumn{11}{c}{$\Pm = 2$}\\
                \hline
                \texttt{weakPm2} & 500 & 1000   & $2\times 10^{-3}$ & $1\times 10^{-3}$ & 0.5 & $1 \leq |\vecB{k}L/2\pi|\leq 3$ & $10^{-10}$    & 0.25 $\pm$ 0.06 & 2.03 $\pm$ 0.25  & $288^3$ \\
                \texttt{satPm2} & 500 & 1000    & $2\times 10^{-3}$ & $1\times 10^{-3}$ & 0.5 & $1 \leq |\vecB{k}L/2\pi|\leq 3$ & $10^{-2}$     & 0.25 $\pm$ 0.05 & 2.05 $\pm$ 0.25  & $288^3$ \\
                \texttt{strongPm2} & 500 & 1000 & $2\times 10^{-3}$ & $1\times 10^{-3}$ & 0.5 & $1 \leq |\vecB{k}L/2\pi|\leq 3$ & $10^{2}$      & 0.26 $\pm$ 0.07 & 2.01 $\pm$ 0.26  & $288^3$ \\
                \texttt{strongPm2\_36} & 500 & 1000 & $2\times 10^{-3}$ & $1\times 10^{-3}$ & 0.5 & $1 \leq |\vecB{k}L/2\pi|\leq 3$ & $10^{2}$  & 0.09 $\pm$ 0.04 & 3.33 $\pm$ 0.74  & $36^3$ \\
                \texttt{strongPm2\_72} & 500 & 1000 & $2\times 10^{-3}$ & $1\times 10^{-3}$ & 0.5 & $1 \leq |\vecB{k}L/2\pi|\leq 3$ & $10^{2}$  & 0.17 $\pm$ 0.06 & 2.42 $\pm$ 0.43  & $72^3$ \\
                \texttt{strongPm2\_144} & 500 & 1000 & $2\times 10^{-3}$ & $1\times 10^{-3}$ & 0.5 & $1 \leq |\vecB{k}L/2\pi|\leq 3$ & $10^{2}$ & 0.24 $\pm$ 0.07 & 2.04 $\pm$ 0.30  & $144^3$ \\
                \texttt{initbPm2} & 500 & 1000  & $2\times 10^{-3}$ & $1\times 10^{-3}$ & 0.5 & $7 \leq |\vecB{k}L/2\pi|\leq 9$ & $10^{2}$      & 0.27 $\pm$ 0.07 & 1.98 $\pm$ 0.25  & $288^3$ \\ 
                \hline
                \multicolumn{11}{c}{$\Pm = 4$}\\
                \hline
                \texttt{weakPm4} & 500 & 2000    & $2\times 10^{-3}$ & $5\times 10^{-4}$ & 0.5 & $1 \leq |\vecB{k}L/2\pi|\leq 3$ & $10^{-10}$  & 0.40 $\pm$ 0.07  & 1.60 $\pm$ 0.15  & $288^3$ \\
                \texttt{satPm4} & 500 & 2000     & $2\times 10^{-3}$ & $5\times 10^{-4}$ & 0.5 & $1 \leq |\vecB{k}L/2\pi|\leq 3$ & $10^{-2}$   & 0.40 $\pm$ 0.08  & 1.60 $\pm$ 0.16  & $288^3$ \\
                \texttt{strongPm4} & 500 & 2000  & $2\times 10^{-3}$ & $5\times 10^{-4}$ & 0.5 & $1 \leq |\vecB{k}L/2\pi|\leq 3$ & $10^{2}$    & 0.41 $\pm$ 0.08  & 1.58 $\pm$ 0.15  & $288^3$ \\
                \texttt{initbPm4} & 500 & 2000   & $2\times 10^{-3}$ & $5\times 10^{-4}$ & 0.5 & $7 \leq |\vecB{k}L/2\pi|\leq 9$ & $10^{2}$    & 0.41 $\pm$ 0.09  & 1.59 $\pm$ 0.18  & $288^3$ \\  
            \hline\hline
    \end{tabular}
    \end{center}
    \begin{tablenotes}[para]
         \textit{\textbf{Notes.}} Column (1): the unique simulation ID. Column (2): the Reynolds number of the plasma, \autoref{eq:Re}, in the saturated phase of the dynamo. Column (3): the same as column (2) but for the magnetic Reynolds number, \autoref{eq:Rm}. Column (4): the coefficient for the kinematic viscosity (see viscous stress tensor in \autoref{eq:momentum}) expressed in units of correlation time of the driving $t_0$ and driving scale $\ell_0$. Column (5): the same as column (4) but for the Ohmic resistivity (\autoref{eq:induction}). Column (6): the turbulent Mach number, $\M = \Exp{\v^2}{\V}^{1/2}/c_s$, where $c_S$ is the sound speed, in the saturated phase of the dynamo. Column (7): the domain of the parabola for the initial magnetic field. Column (8): the initial magnetic and kinetic energy ratio. Column (9): the magnetic and kinetic energy ratio in the saturated state of the dynamo. Column (10): the same column (9) but for the Alfv\'en Mach number, $\Ma = \Exp{v^2}{\V}^{1/2}/\Exp{v_A^2}{\V}^{1/2}$. Column (11): the grid resolution of the simulation.
    \end{tablenotes}
\end{table*}

    In this first of two ``Growth or Decay" studies on the turbulent dynamo, we establish that the dynamo saturation is universal for both initial magnetic field structure and amplitude, at least for moderate magnetic Reynolds numbers, comparable to those found in the laboratory (e.g., $\Rm \approx 600$, \citealt{Tzeferacos2018_dynamo_lab_experiment_nature}; $\Rm =450\pm220$, \citealt{Bott2021_time_resolved_dynamo}). Specifically, the $t \rightarrow \infty$ integral quantities, energy spectra, and characteristic scales of the magnetic field do not depend upon the initial conditions of $\vecB{b}$, as hypothesised by \citet{Maron2004_nonlinear_magnetic_spectrum}, but not explored systematically. This means that the kinematic and non-linear stages of the dynamo do not produce a set of initial conditions that uniquely define the structure of the magnetic field in the saturated state. Hence, the saturated field in, for example, the ISM, but also other turbulent plasma phenomena, such as the accretion disk of black holes \citep[e.g.,][]{Ripperda2020_black_hole_accretion_disk}, ought to be determined by the turbulence\footnote{Assuming that the magnetic dissipative mechanisms (based on the microphysics of the gas) are somewhat universal on small scales.}. 
        
    \subsection{Organisation of our study}  
    Our study is organised as follows. In \autoref{sec:sims} we discuss the numerical simulations that we use to probe how the saturation of the SSD responds to changing the initial ratio between the magnetic and kinetic energy and the scales that the magnetic energy are initialised upon. In \autoref{sec:integral_quants} we report upon the ratio of integral energies. Next in \autoref{sec:spectrum} we explore the saturation on a scale-by-scale manner, exploring the time-dependent energy spectra ratios as the dynamo progresses towards saturation, followed by \autoref{sec:magnetic_field_scales}, where we study the characteristic scales -- integral, peak energy and micro scales -- of the magnetic energy. In \autoref{sec:implications_and_limitations} we briefly discuss the implications of the dynamo saturation being universal and the limitations of our study. Finally, in \autoref{sec:conclusion} we summarise and list the key results of our study.

\section{Numerical simulations}\label{sec:sims}
        \subsection{Fluid model and numerical code}
        We use a modified version of the finite volume \textsc{flash} code \citep{Fryxell2000,Dubey2008}, utilising a second-order conservative MUSCL-Hancock 5-wave approximate Riemann scheme \citep{Bouchut2010,Waagan2011,Federrath2021}, utilising $\nabla \cdot \vecB{b}$ parabolic diffusion flux cleaning \citep{Marder1987_fluxcleaning} to solve the 3D, visco-resistive, isothermal, compressible MHD equations with a stochastic non-helical acceleration field acting to drive the turbulence,
        \begin{align}
            \frac{\partial \rho}{\partial t} + \nabla\cdot(\rho \vecB{v}) &= 0 \label{eq:continuity}, \\
            \frac{\partial\rho\vecB{v}}{\partial t}  - \nabla\cdot\left[ \frac{1}{4\pi}\vecB{b}\otimes\vecB{b} - \rho \vecB{v}\otimes\vecB{v} \right.\nquad[5]& \nonumber\\
            \left. - \left(c_s^2 \rho + \frac{b^2}{8\pi}\right)\mathbb{I} + 2\nu\rho\mathbb{S}\right] &= \rho \vecB{f},\label{eq:momentum} \\
            \frac{\partial \vecB{b}}{\partial t} - \nabla \times\left[\vecB{v} \times \vecB{b} - \eta\vecB{j}\right]&= 0,\label{eq:induction}\\
            \nabla \cdot \vecB{b} &= 0, \label{eq:div0}
        \end{align}
        where $\mathbb{I}$ is the unit tensor. We solve the equations on a periodic domain of dimension $L^3\equiv \V$, with $288^3$ grid cells, where $\vecB{v}$ is the fluid velocity, $\rho$ is the gas density, $\vecB{j} = (\nabla \times \vecB{b})/(4\pi)$ is the current density, $\vecB{b}$ is the magnetic field, where $\Exp{\vecB{b}(t)}{\V}=0$, $c_s$ is the sound speed, $\mathbb{S}$ is the strain rate tensor, $\mathbb{S} = (1/2)(\nabla \otimes \vecB{v} + [\nabla \otimes \vecB{v}]^T) - (1/3)(\nabla\cdot\vecB{v})\mathbb{I}$\footnote{\label{fn:incompressible} Note that for an incompressible fluid, $\nabla\cdot\vecB{v} = 0$, $\rho = \rho_0 = \text{const.}$, and then $\nabla\cdot\nu\mathbb{S} = \nu\nabla^2\vecB{v}$, as expected.}, and $\vecB{f}$, the stochastic turbulent acceleration source term that drives the turbulence. In the ISM, $\vecB{f}$ could be from, for example, supernova shocks, internal instabilities in the gas, gravity, galactic-scale shocks and shear, or ambient pressure from the galactic environment \citep{Brunt2009,Elmegreen2009IAUS,Federrath2015_inefficient_SFR,Krumholz2016,Grisdale2017,Jin2017,Kortgen2017,Federrath2017IAUS,Krumholz2018_unified_disc_model,Colling2018,Schruba2019,Lu2020_supernova_driving}. Both the viscosity $\nu$ and resistivity $\eta$ coefficients are constant in space and time. We perform a set of convergence tests for the saturation of the magnetic field in \aref{appendix:convergence}. 
    
    \subsection{Turbulent driving}\label{sec:turb_driving}
        The forcing term $\vecB{f}$ follows an Ornstein-Uhlenbeck process with finite $e$-fold correlation time, $t_0 = \lo/\Exp{\v^2}{\V}^{1/2}$. $\vecB{f}$ is constructed in Fourier space with energy injected on the peak scale $|\vecB{k}L/2\pi|=2$ (equivalently, $\lo = L/2$) and falls off to zero with a parabolic spectrum within $1 \leq |\vecB{k}L/2\pi| \leq 3$. On $\lo$, we use the correlation time and Fourier amplitude to control the rms velocity, which we set to $\Exp{\v^2}{\V}^{1/2}/c_s = \M = 0.5$ in the saturated stage of the dynamo, which, up to $2\Exp{\v^2}{\V}^{1/2}$ in the velocity distribution, corresponds to an incompressible flow. We inject energy isotropically and solely into the solenoidal ($\nabla \cdot \vecB{f}=0$) mode component of $\vecB{f}$ \citep[as solenoidal driving gives a higher dynamo efficiency in comparison to compressive driving, see ][]{Federrath2011_mach_dynamo,Chirakkara2021}. See \citet{Federrath2008,Federrath2009,Federrath2010,Federrath2022_turbulence_driving_module} for more details about the turbulent driving. We run each of the experiments from $t/t_0=0$ to $t/t_0=1000$, writing the 3D field variables to disk every $t/t_0=0.5$ to ensure we produce a dataset that resolves each of the dynamo stages in time, and is able to capture well-sampled statistics from the magnetic field in the large $t$ limit. 
        
    \subsection{Dimensionless plasma numbers}
        Apart from $\M$, there are three main dimensionless numbers that we use to both parameterise and contextualise the results of our simulations. The first is the hydrodynamic Reynolds number,
        \begin{align}\label{eq:Re}
            \Re = \frac{|\nabla\cdot(\vecB{v}\otimes\vecB{v})|}{|\nu\nabla^2\vecB{v}|} \sim \frac{\Exp{\v^2}{\V}^{1/2}\lo}{\nu},
        \end{align}
        which informs us of the relative strength for the Reynolds stress $|\nabla\cdot(\vecB{v}\otimes\vecB{v})|$ compared with the dissipation $|\nu\nabla^2\vecB{v}|$ (assuming incompressibility; see \autoref{fn:incompressible}) in \autoref{eq:momentum} (the momentum equation). This number also provides a measure of the width for the range of scales that are self-similar in the turbulence -- part of the non-linear turbulent cascade, i.e., $\lnu \ll \ell \ll \lo$, where $\lnu \sim \Re^{3/4}\lo$. By setting $\nu$, for a fixed $\Exp{\v^2}{\V}^{1/2}\lo$, we are able to control $\Re$ for each of our simulations. In this study we use $\Re = 500$ for all of our simulations, as indicated in column (2) of \autoref{tab:sims}.
        
        The second dimensionless parameter in our simulations is the magnetic Reynolds number,
        \begin{align}\label{eq:Rm}
            \Rm = \frac{|\nabla\times(\vecB{v}\times\vecB{b})|}{|\eta\nabla\times\vecB{j}|} \sim \frac{\Exp{\v^2}{\V}^{1/2}\lo}{\eta},
        \end{align}
        which is analogous to $\Re$, and compares the induction $|\nabla\times(\vecB{v}\times\vecB{b})|$ and dissipation $|\eta\nabla\times\vecB{j}|$ terms in \autoref{eq:induction} (the induction equation). By setting $\eta$ we control $\Rm$ and vary it between 500 and 2000, as indicated in column (3) of \autoref{tab:sims}, ensuring that we are significantly above the critical $\Rm$ for the SSD to take place ($\Rm \sim 100$; \citealt{Ruzmaikin1981_rm_crit,Haugen2004_dynamo_mach_and_crit_rm,Schekochihin2004_critical_dynamo,Federrath2014_supersonic_dynamo,Seta2020_saturation_high_Pm}).
        
        The final is the Prandtl number, which is simply the ratio between the two plasma Reynolds numbers,
        \begin{align}
            \Pm = \frac{\nu}{\eta} \sim \frac{\Rm}{\Re}.
        \end{align}
        In units of the correlation time of the forcing\footnote{Note $[\nu] = [\eta] \sim \ell^2/t_{\rm diffuse}$, and that $t_{\eta}/t_0 = (\ell_\eta^2/\ell_0)\eta^{-1}\Exp{\v^2}{}^{1/2}$, which is $t_{\eta}/t_0 = \Rm^{-1}$ for $\ell_\eta \sim \ell_0$.}, $\Pm$ is $\sim$ the ratio between diffusion timescales for the magnetic $t_{\eta}$ and kinetic $t_{\nu}$ fluctuations, $\Pm = t_{\eta}/t_{\nu}$. For $\Pm = 1$, the diffusion timescales are equal, and thus it naturally follows that $\leta \sim \lnu$ where $\leta$ is the magnetic dissipation scale. Equivalently, $\Pm$ also provides a measure of the scale separation between $\lnu$ and $\leta$. For \citet{Kolmogorov1941} turbulence in a magnetised plasma with $\Pm \gg 1$, as is the case for the ISM, $\lnu / \leta \gg 1$, and $\lnu / \leta \sim \Pm^{1/2}$ (\citealt{Schekochihin2002_large_Pm_dynamos} derived this relation by balancing viscous stretching with magnetic dissipation, which was recently confirmed by \citealt{Kriel2022_kinematic_dynamo_scales} and \citealt{Brandenburg2022_SSD_scales_dissipation} using direct numerical simulations). Similarly, for $\Pm \ll 1$ plasmas, $\lnu / \leta \ll 1$, and $\lnu / \leta \sim \Pm^{3/4}  \ll 1$ \citep{moffatt1961_low_Pm_dynamos} as is the case for liquid metal experiments, stars, and planetary plasmas \citep{Rincon2019_dynamo_theories}. In this study we will be focusing on $\Pm \geq 1$ plasmas, relevant to the ISM, albeit without being able to venture very far from $\Pm = 1$ (varying $\Pm = 1-4$) due to the limited simulation grid resolution available to us. 

        \begin{figure*}
            \centering
            \includegraphics[width=\linewidth]{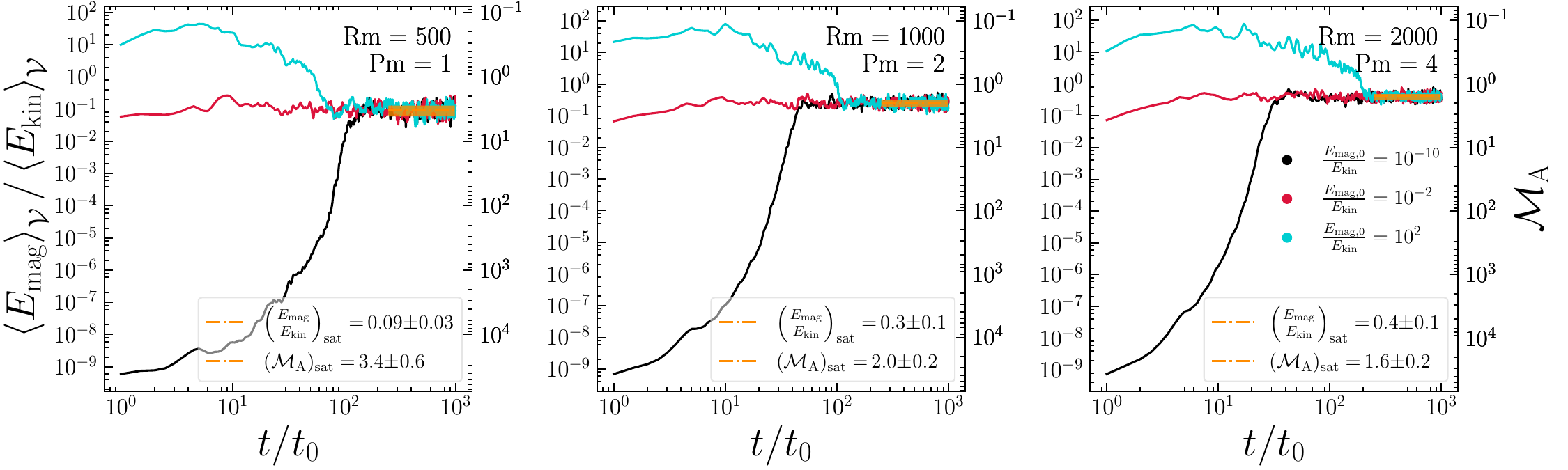}
            \caption{Time-dependent integral energy ratios (left axis) and $\Ma$ (right axis) for the $\Rm=500$, $\Rm=1000$ and $\Rm=2000$ simulations with $b_{\text{init}}=1 \leq |\vecB{k}L/2\pi|\leq 3$ (see \autoref{tab:sims}). Each column corresponds to a different $\Rm$ ensemble, indicated by the annotation above the column. Different initial magnetic field strengths are indicated by the colouring, shown in the legend of the rightmost panel. Regardless of the initial magnetic field strengths, the turbulent dynamo produces a universal saturation value for each $\Rm$ simulation ensemble, which we annotate in the bottom right corner of each panel.}
            \label{fig:Rm_energy_balance}
        \end{figure*} 

        \begin{figure}
            \centering
            \includegraphics[width=\linewidth]{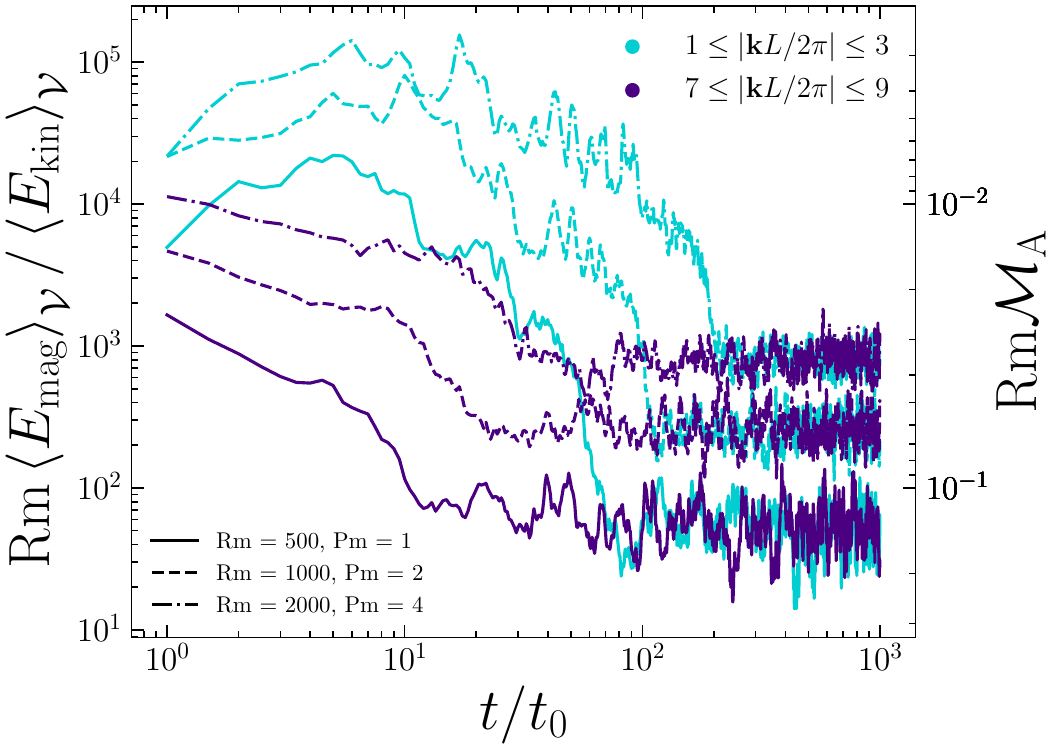}
            \caption{$\emagb/\ekinb$ (left) and $\Ma$ (right) as a function of $t/t_0$ for different initial magnetic field structure. All experiments are scaled by $\Rm$ to separate the curves and clearly show the behaviour of the energy ratios as $t/t_0 \rightarrow \infty$. The \texttt{strong} experiments from the main study are shown in aqua and experiments with $\b_{\rm init} = 7 \leq |\vecB{k}/2\pi|\leq 9$, the \texttt{init} experiments, are shown in purple.}
            \label{fig:init_b_field}
        \end{figure}

    \subsection{Initial conditions}
        The initial velocity field in our simulations is set to $|\vecB{v}(x,y,z,t=0)|/c_s=0$, with units $c_s=1$, and the density field is initialised to a constant value, $\rho(x,y,z,t=0)=\rho_0$, where the density has units $\rho_0=1$. We do not initialise our simulations with a mean-field component of the magnetic field, $\Exp{\vecB{b}}{\V} = 0$, instead including only a fluctuating-component of the magnetic field, $\vecB{b}(x,y,z,t=0)/(c_s\rho_0^{1/2})$, which has units $c_s\rho_0^{1/2} = 1$. This is an obvious and useful non-dimensionalisation of \autoref{eq:continuity}--\ref{eq:div0} for an isothermal fluid, which allows us to scale to an arbitrary dimensionalisation of an isothermal systems, for example, any approximately isothermal phase of the interstellar gas \citep{Wolfire1995_isothermal_ISM}.
        
        To explore the universality of the dynamo saturation, we set the initial magnetic field $\vecB{b}(t=0)$ to be in one of four configurations for each $\Pm$ experiment. In three of the four configurations we set $\vecB{b}(t=0)$ with the same $\b_{\text{init}} \equiv 1 \leq |\vecB{k}L/2\pi|\leq 3$ parabolic field as the driving momentum field (see, \autoref{sec:turb_driving}) but with $E_{\text{mag},0}/\ekin = \left\{10^{-10}, 10^{-2}, 10^2\right\}$ (column (8) in \autoref{tab:sims}), where
        \begin{align}
        \emagb \equiv \frac{\Exp{\b^2}{\V}}{8\pi} = \frac{L}{16\pi^2} \int_{0}^{\infty}\d{k}\,\Exp{|\vecB{b}(\vecB{k})|^2}{\theta},
        \end{align}
        is the integral magnetic energy, and $E_{\text{mag},0}$ is the magnetic energy at $t=0$. The integral kinetic energy is likewise defined as 
        \begin{align}   
        \ekinb \equiv \frac{\rho_0\Exp{\v^2}{\V}}{2} = \frac{\rho_0 L}{4\pi} \int_{0}^{\infty}\d{k}\,\Exp{|\vecB{v}(\vecB{k})|^2}{\theta},
        \end{align}  
        where $\Exp{|\vecB{b}(\vecB{k})|^2}{\theta}$ and $\Exp{|\vecB{v}(\vecB{k})|^2}{\theta}$ are the 1D shell-integrated (over $\theta$) power spectra. The first (which we call the \texttt{weak} experiments) of the three is a classical dynamo experiment, which leads to $\emag$ evolving through all three of the dynamo stages: kinematic, nonlinear, and saturation. The second (\texttt{sat} experiments) are initialised such that $E_{\text{mag},0}/\ekin \approx (\emag/\ekin)_{\rm sat}$, where $(\emag/\ekin)_{\rm sat}$ is the saturated state of the energy ratio measured from the \texttt{weak} experiment. Next, the third (\texttt{strong} experiments) are initialised with roughly four orders of magnitude more $\emagb$ than is supported by the SSD. The final configuration is also a \texttt{strong} experiment, but $\vecB{b}(t=0)$ is initialised on smaller scales than in the other experiments, also with an initial parabolic profile, but peaking at $|\vecB{k}L/2\pi|=8$ (equivalently, $\ell = L/8$) and falling off within $\b_{\rm init} = 7 \leq |\vecB{k}L/2\pi| \leq 9$. We call these experiments \texttt{initb} and compare them to the other \texttt{strong} experiments. The final configuration allows us to probe how the strong-field experiments respond to having the bulk of the $\emagb$ initialised on high-$k$ modes, close to scales that ought to be dominated by dissipation. 

        \begin{figure*}
            \centering
            \includegraphics[width=\linewidth]{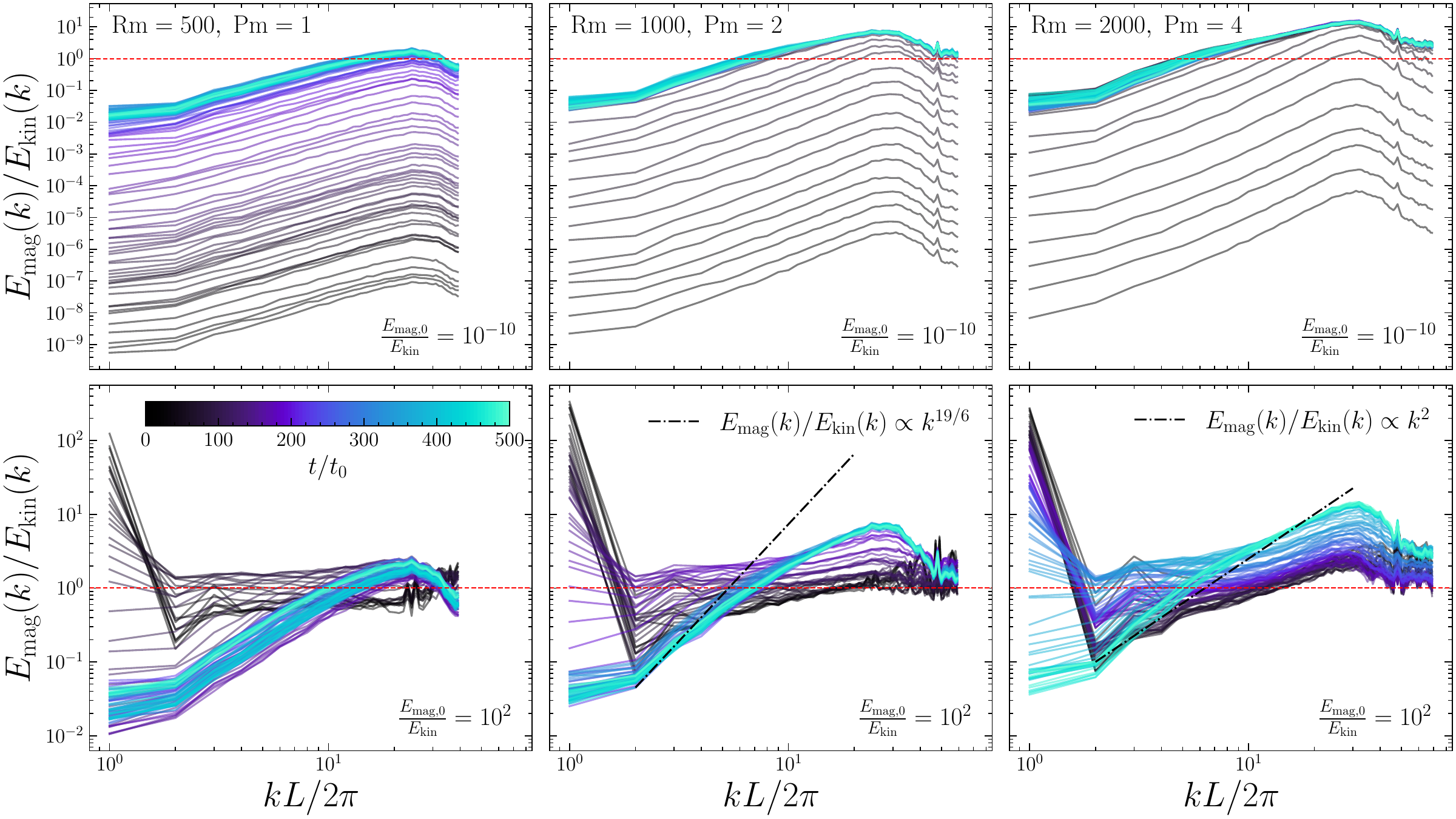}
            \caption{\textbf{Top:} the temporal evolution (black, $t/t_0 \sim 0$; aqua, $t/t_0 \sim 500$) of the magnetic energy as a function of $k$-modes $\emag(k)$, normalised by the kinetic energy $\ekin(k)$ for the classical small-scale dynamo experiment, (\texttt{weak}). The red horizontal line indicates energy equipartition. \textbf{Bottom:} The same as the top panel, except for the strong-field (\texttt{strong}) decaying dynamo experiment. In the decaying regime, the low-$k$ modes lose energy slowly (only settling after $t\gtrsim100t_0$). However, simultaneous to the decay, the high-$k$ modes are being amplified (or maintained for $\Pm=1$, on roughly the same timescale as the low-$k$ mode decay) by the turbulence. When the low-$k$ modes have decayed, and the high-$k$ modes have been amplified beyond energy equipartition and the saturated state of the small-scale dynamo is reached. For all experiments, the ratios of the spectra exhibit an extended self-similar structure, $\emag(k)/\ekin(k)\propto k^2$ (shown in bottom-right plot). We also show $\emag(k)/\ekin(k) \propto k^{3/2}/k^{-5/3} = k^{19/6}$ (bottom-middle panel) for \citet{Kazantsev1968} magnetic and \citet{Kolmogorov1941} velocity spectrum. We show a similar plot but for just the $\ekin(k)$ in \autoref{fig:kinetic_spectra} and just $\emag(k)$ in \autoref{fig:magnetic_spectra}. All spectral ratios are truncated at the scales dominated by numerical dissipation to highlight only the ratios within the resolved modes.} 
        \label{fig:time_evolution_spectra}
        \end{figure*}

        \begin{figure}
            \centering
            \includegraphics[width=\linewidth]{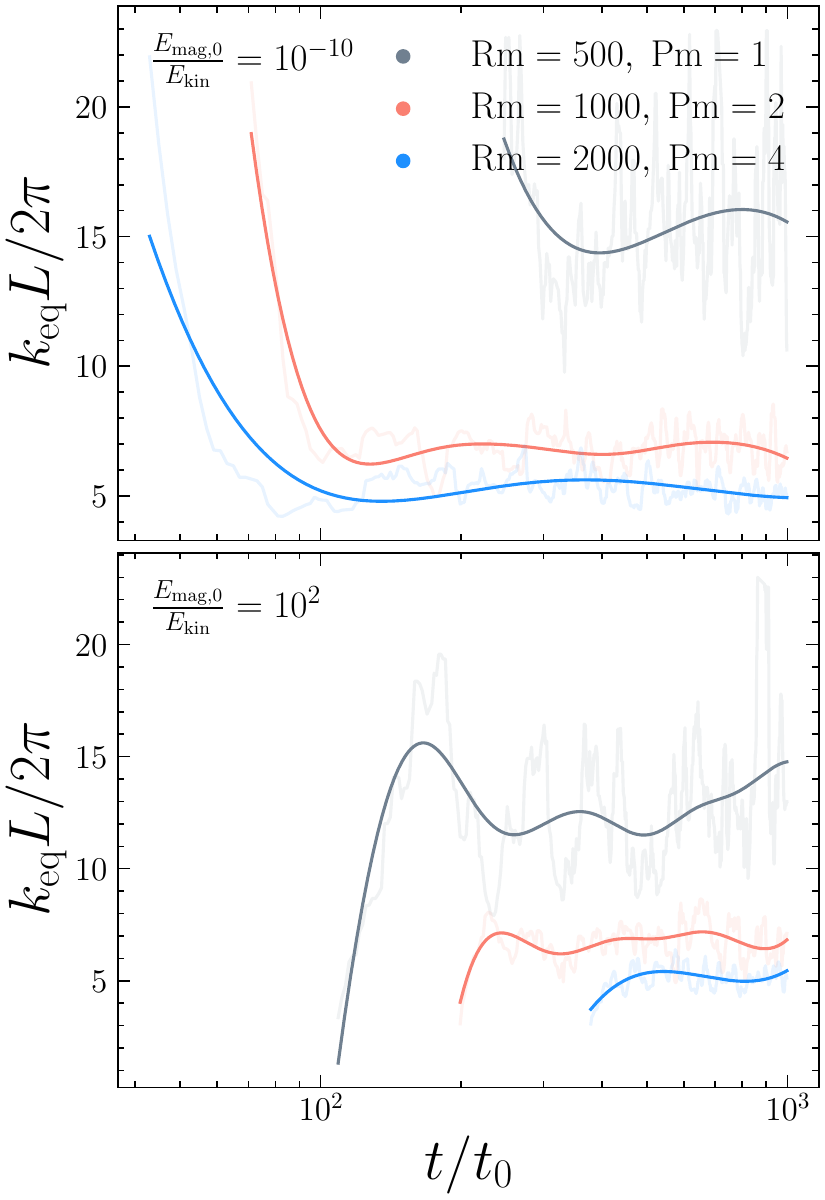}
            \caption{The energy equipartition scale, $k_{\rm eq}: \emag(k) = \ekin(k)$ as a function of $t/t_0$, coloured by $\Pm$. \textbf{Top:} the classical dynamo experiment, \texttt{weak}, growing into the saturated stage, showing  a scale-by-scale saturation in effect, starting at high-$k$ modes and moving towards low $k$-modes. \textbf{Bottom:} the experiment initialised with a strong magnetic field \texttt{strong}, decaying into the saturated stage, showing the opposite scale-by-scale saturation -- low-$k$ modes to high-$k$ modes.}
            \label{fig:k_eq}
        \end{figure}

        \begin{figure*}
            \centering
            \includegraphics[width=\linewidth]{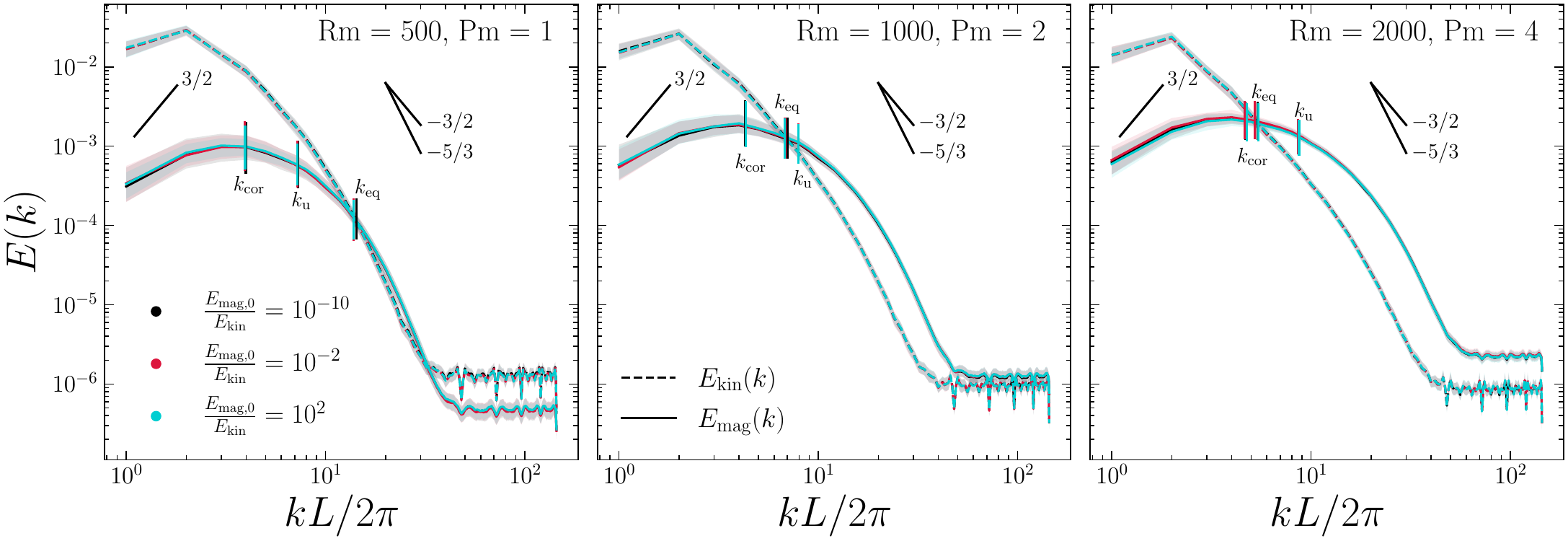}
            \caption{Time-averaged energy spectra in the saturated dynamo state, coloured and organised by $\Rm-\Pm$ as in \autoref{fig:Rm_energy_balance}, with different linestyles for $\ekin(k)$ and $\emag(k)$, as indicated by the legend in the middle panel. Magnetic turbulent wavenumbers directly computed from the $\emag(k)$ are annotated in each panel, where $\kcor$ is the correlation wavenumber (\autoref{eq:correlation_scale}; the scale of the largest correlated turbulent structures), $\ku$ is the micro or rms scale (\autoref{eq:micro_scale}; the scale of the smallest correlated turbulent structures) and $\keq$ is the energy equipartition scale (\autoref{eq:k_eq}; the lowest wavenumber where $\emag(k)=\ekin(k)$). The \citet{Kazantsev1968} ($k^{3/2}$; top-left), \citet{Kolmogorov1941} and \citet{Boldyrev2006} ($k^{-5/3}$ and $k^{-3/2}$, respectively; top-right) spectra are annotated in black.}
            \label{fig:averged_ps}
        \end{figure*}

\section{Integral energy quantities}\label{sec:integral_quants}

    \subsection{Influence of changing the initial magnetic energy}
        In \autoref{fig:Rm_energy_balance} we plot the integral energy ratio $\emagb/\ekinb$ (left axis) and $\Ma = (\emagb/\ekinb)^{-2}$ (right axis) as a function of time, in units of the correlation times of the turbulence driving, $t/t_0$. In each panel we show a different $\Rm$ (labelled in the top right; and hence a different $\Pm$), showing the most resistive simulations on the left and the least on the right. Each colour represents a different $E_{\text{mag},0}/\ekin$, indicated in the legend on the rightmost panel (\texttt{weak}, black; \texttt{sat}, red; \texttt{strong}, aqua). This colour scheme will be maintained throughout the remainder of the study. 
        
        For each $\Rm$ (panel), regardless of $E_{\text{mag},0}/\ekin$, the integral energies for each of the experiments reach the same saturated value, which varies between roughly $10\%$ and $40\%$ in $\emagb/\ekinb$ (corresponding to $\Ma = 3.4 - 1.6$, respectively), labelled on each plot with the orange line (band showing 1$\sigma$ within the averaging range) and clearly depends upon $\Rm$. We list all the saturated values of the energy ratio and $\Ma$ in \autoref{tab:sims}. The journey towards the saturation is quite different between the three different $E_{\text{mag},0}/\ekin$. The \texttt{weak} runs explore the full three stages of the SSD, whilst the \texttt{sat} experiments are immediately in, and maintained at the saturated state, at least according to $\emagb/\ekinb$. Of course, the magnetic field may be being reorganised by the turbulence in $k$-space, but regardless, $\emagb/\ekinb$ remains constant at the saturated level. For the first $10t_0$ the \texttt{strong} experiments remain strong, and in fact $\emagb/\ekinb$ increases by roughly an order of magnitude (discussed further in the following section). But these dynamics are short-lived, and beyond $10t_0$ $\emagb/\ekinb$ is clearly decaying via a two-stage exponential into the universal $t\rightarrow\infty$ state. The first exponential process is slow, and the second is significantly faster, which is the opposite of the kinematic and linear growth stages in the classical dynamo experiment. In Paper II, we will explore the timescale of the decay in much more detail, which is clearly a function of $\Rm$, with the most resistive (lowest $\Rm$) experiments taking shorter $t/t_0$ paths towards the saturation. The key result is that the final state of  $\emagb/\ekinb$ does not depend upon $E_{\text{mag},0}/\ekin$.

    \subsection{Influence of changing initial magnetic field structure}\label{appendix:init_mag_field}        
        Performing a similar analysis as \citet{Seta2020_seed_magnetic_field}, but for the \texttt{strong} decaying experiments, we show in \autoref{fig:init_b_field} the same energy ratio plot as in \autoref{fig:Rm_energy_balance} but for two sets of experiments where $\b_{\rm init} = 1 \leq |\vecB{k}/2\pi|\leq 3$ (\texttt{strong}, aqua; the simulations from the main study) and $\b_{\rm init} = 7 \leq |\vecB{k}/2\pi|\leq 9$ (\texttt{init}, purple; initialised as a small-scale field). Energy ratios are scaled by $\Rm$ to separate the simulations at $t/t_0 \gg 1$. Like \citet{Seta2020_seed_magnetic_field}, we show that regardless of the initial magnetic field structure, $\emagb/\ekinb$ and $\Ma$ remain the same for each different $\Rm$ in saturation. The \texttt{strong} simulations take a longer time to decay compared to the \texttt{init} simulations, most likely because the large-scale field in the \texttt{strong} simulations takes a long time to be destroyed in this $\Rm$ regime, since both the nonlinear and dissipation timescales are shrinking as we move to smaller scales. 

        Moving beyond the integral energy quantities of the saturation, we now turn to the scale-by-scale representation of the same ratio plots in $k$-space, with focus on the experiments with $\b_{\rm init} = 1 \leq |\vecB{k}/2\pi|\leq 3$ for the remainder of the study.
        
\section{Energy spectra}\label{sec:spectrum}
    \subsection{The two different journeys towards saturation}
    In \autoref{fig:time_evolution_spectra} we show the ratio between the time-dependent 1D magnetic $\emag(k)=\Exp{|\vecB{b}(\vecB{k})|^2}{\theta}/(8\pi)$ and kinetic $\ekin(k)=\rho_0\Exp{|\vecB{v}(\vecB{k})|^2}{\theta}/2$ energy spectrum for the \texttt{weak} experiment (top panel) and \texttt{strong} experiment (bottom panel). Each spectrum is coloured by $t/t_0$, varying from black $t/t_0 \sim 0$ to light aqua $t/t_0 = 500$. In each column we show each $\Rm$-$\Pm$ combination, which are all in the saturated state by $500t_0$ (see \autoref{fig:Rm_energy_balance}) -- the maximum correlation time we plot in this figure. We annotate the energy equipartition $\emag(k) = \ekin(k)$ with the red-dashed line in both panels, and directly plot the time-evolution of $\keq: \emag(\keq) = \ekin(\keq)$ (see \autoref{appendix:k_eq} for details on how we define this scale, which is consistent with the saturation phenomenology we described in \autoref{sec:intro}) for the \texttt{weak} (top) and \texttt{strong} (bottom) panels in \autoref{fig:k_eq}, coloured by $\Rm$ and $\Pm$. 
        
    The top panels of \autoref{fig:time_evolution_spectra} correspond to the classical SSD experiment, which is where we will begin our analysis. Firstly, within a few $t_0$, $\emag(k)/\ekin(k)$ is quickly organised into a self-similar state, peaked on high-$k$ modes -- scales close to (if not at) the resistive scale $\keta$. This ratio is maintained through the whole kinematic stage (the black curves) but flattens at $k$ above $\keq$ as the plasma approaches saturation (light aqua curves). We will discuss this morphology in more detail when focusing on the \texttt{strong} experiments. $\d{\emag(k)}/\d{(t/t_0)}$ can be observed \footnote{Note that all spectra are sampled at the same rate, $\sim 2/t_0$, hence ``large" differences (see the kinematic stage in the top-right $\Pm=4$ panel) between any successive spectra correspond to ``large" $\d{\emag(k)}/\d{(t/t_0)}$, and likewise for small differences (see low-$k$ modes decaying in the bottom-right $\Pm=4$ panel).}, and seems to be approximately constant across all $k$ in the kinematic stage (the turbulence grows all modes evenly once the field has been reorganised).
    
    $\keq$ (the $k$ scale intersecting the red-dashed line) in \autoref{fig:k_eq}, shows that as the experiments (top panel) approach saturation, $\keq$ moves towards larger scales (smaller $k$), as eddies on successively larger scales become responsible for amplifying the magnetic field \citep{Galishnikova2022_saturation_and_tearing}. However, as found previously in, e.g., \citet{Maron2004_nonlinear_magnetic_spectrum}, an exact scale-by-scale energy equipartition is not realised, and in the saturated stage, the magnetic field is able to be maintained by the turbulence in superequipartition on scales where $\emag(k>\keq) > \ekin(k>\keq)$. This means that the timescale for the coupling between the kinetic and magnetic energy $t_{\rm couple} \sim \big(\Exp{\v^2}{\V}^{1/2}/L\big)^{-1}\big(\Exp{\v^2}{\V}^{1/2}/\Exp{\b^2}{\V}^{1/2}\big)^{-2}$ is shorter than the magnetic energy diffusion timescale $t_{\eta} \sim \eta^{-1} \ell^{2}$ on these scales (assuming that these scales are dominated by diffusion) $t_{\rm couple} < t_{\eta}$, i.e., there are multiple across-field (not necessarily local) coupling events, feeding and growing the magnetic energy, per events that are able to dissipate it via Ohmic dissipation (or other means), consistent with the transfer function analysis performed in \citet{Galishnikova2022_saturation_and_tearing} (they frame this phenomenon as the injection energy doing work against the Lorentz force). This is exacerbated as $\Pm$ increases, which increases $t_{\eta}$, and shifts $\keq$ to lower $k$-modes. Now we turn our attention to the bottom panel of \autoref{fig:time_evolution_spectra}.
        
    In the bottom panels of \autoref{fig:time_evolution_spectra} we observe a different pathway to the same saturated $\emag(k)$ state, similarly to what \citet{Maron2004_nonlinear_magnetic_spectrum} discussed in \S5.4 of their study. Unlike the \texttt{weak} experiment, $\d{\emag(k)}/\d{(t/t_0)}$ is very different on large and small scales in the \texttt{strong} experiments. Due to the initialisation of the $\vecB{b}$-field, $\emag(k)$ is concentrated on the largest scales for $t/t_0 \sim 0$. As is demonstrated from the top panel in the \texttt{weak} experiment, these modes are not able to be maintained by the SSD, and start decaying slowly until $\emag(k)\sim\ekin(k)$, and then quickly into the final saturated state where $\emag(k)<10^{-1}\ekin(k)$ (the exact values depend upon $\Rm$, even at these low modes). This final fast decay stage can be seen in the integral energy plots, \autoref{fig:Rm_energy_balance}, where $\emagb/\ekinb$ crashes sharply before being maintained in the saturated stage. In contrast, as the low-$k$ magnetic modes decay, $\keq$ moves to smaller scales (the opposite as the \texttt{weak} experiments; bottom panel in \autoref{fig:k_eq}) and the high-$k$ modes grow into the somewhat self-similar, peaked $\emag(k)/\ekin(k)$ structure that we observed in the saturated state for the \texttt{weak} experiments. Because the low-$k$ modes begin their decay slowly, the growth of the high-$k$ modes boosts the integral energy by some amount for small $t \lesssim 10t_0$, as we saw in \autoref{fig:Rm_energy_balance} (at $t\lesssim10t_0$ the $\emagb$ initially grows by an order of magnitude). But even with these modes growing, the amount of energy being lost through the decay of low-$k$ modes surpasses the growth and gives rise to the exponential decay that we observed in \autoref{fig:Rm_energy_balance}. To summarise, the magnetic field initially undergoes a slow decay, but then after some characteristic time undergoes a fast decay into the saturated state. Modelling this process will be a key focus of Paper II. We show the separate energy spectra (not the ratio) in \aref{appendix:kinetic_spectra}.
        
    For both the \texttt{strong} and \texttt{weak} experiments, $\emag(k)/\ekin(k)$ shows a broken power-law structure that roughly extends from $k_0$ to $k_{\text{eq}}$, and then from  $k_{\text{eq}}$ to $k_{\text{max}} = \text{argmax}_{k}\left\{ \emag(k)/\ekin(k)\right\}$, which is reminiscent of an extended self-similarity -- fractal structure in the turbulence encapsulating much more than just the scales that exist in the cascade \citep[inertial range for incompressible turbulence, ][]{Benzi1993_extended_self_similiarity}\footnote{Note that extended self-similarity is classically invoked for measuring extended power-law scalings in velocity structure functions when each order is normalised by the $3^{\rm rd}$-order velocity structure function. This means, to strengthen this analogy, one may seek to construct higher-order statistics of the ratios between the magnetic and kinetic energy, which we leave for future studies to pursue.}.
    We provide rough estimates of these power-laws. In the bottom-central panel, we plot $\emag(k)/\ekin(k)\propto k^{19/6} = k^{3/2}/k^{-5/3}$, for a \citet{Kazantsev1968} magnetic field (a kinematic stage theory) and \citet{Kolmogorov1941} velocity field, but this is not preferred, and instead a $\emag(k)/\ekin(k)\propto k^{2}$ describes the data well over a broad range of $k$, regardless of $\Rm$. However, the measured saturation as a function of scale is the integral of this ratio $\emagb/\ekinb \propto \int_0^{k}\d{k'}\,\emag(k')/\ekin(k') \propto k^3$, adding up all $k$ contributing to the saturation on each scale. This shows that the turbulence has a strong preference to become magnetised on small scales, highlighting the small-scale (compared to kinetic energies) nature of magnetic field energies.
        
    We expect that this narrative does not change in general with $\Rm$ (for a fixed $\Re$), however, the integral energies of the saturated state will change with increasing $\Pm$ because as $\keq$ shifts towards lower-$k$ scales, more and more scales become magnetically dominated, increasing $\emagb/\ekinb$). Next, we look at the average energy spectra in the saturated state for all $\Rm$ explored in this study.
    
    \subsection{Saturated energy spectra}
        We show the kinetic (dashed) and magnetic (solid) energy spectra averaged in the saturated regime in \autoref{fig:averged_ps}, with the same colouring scheme as in \autoref{fig:Rm_energy_balance}, for each $\Rm$, increasing from left-to-right in each of the panels. We label the \citet{Boldyrev2006} (dynamically aligned turbulence\footnote{Or \citet{Kraichnan1965_IKturb} turbulence, i.e., \citet{Kolmogorov1941} with an irreducible magnetic field.}; $k^{-3/2}$) and \citet{Kolmogorov1941} (homogeneous, isotropic turbulence; $k^{-5/3}$) kinetic energy spectra scalings in the top-right of each panel, and the \citet{Kazantsev1968} spectrum for the magnetic field (field folding in the kinematic stage; $k^{3/2}$) in the top-left.
        
        As we have shown in the previous section, but now more clearly for different $\Rm$, the overall shape of the $\emag$ spectra does not depend upon whether the saturated state is reached from below via the two stages of the SSD, or from above through the simultaneous decay and growth of $k$ modes -- that is, at a $k$ mode by $k$ comparison, the \texttt{weak} and \texttt{strong} experiments are practically identical after enough time has passed in the simulations. We can explore the similarities and differences more qualitatively by turning to the characteristic scales of the magnetic field, directly computed from the energy spectrum, which we do now.

    \begin{figure}
        \centering
        \includegraphics[width=\linewidth]{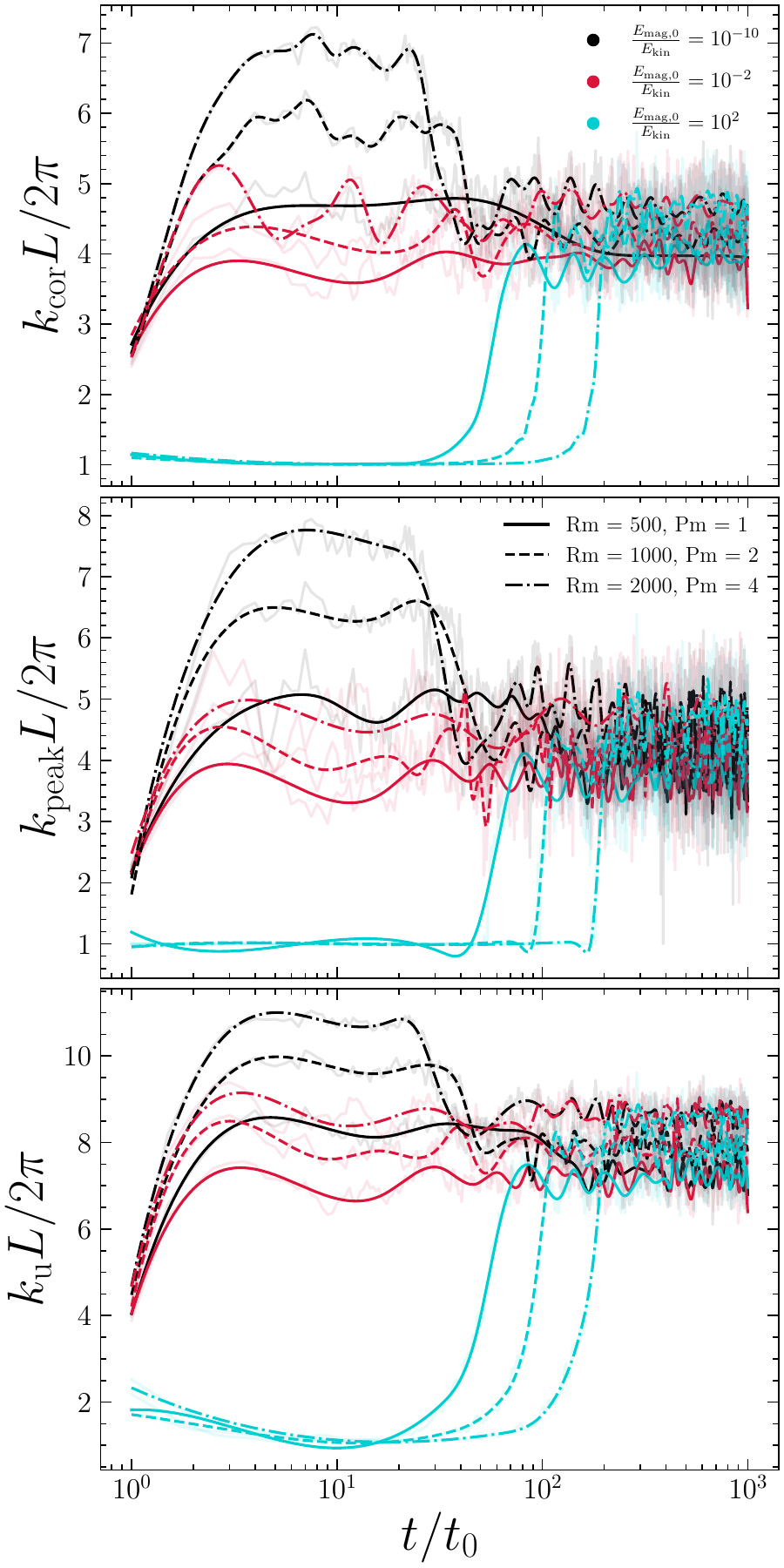}
        \caption{The characteristic magnetic energy spectra wavenumbers, correlation $\kcor$ (\autoref{eq:correlation_scale}), peak scale $\kp$ (\autoref{eq:peak_scale}) and micro $\ku$ (\autoref{eq:micro_scale}), top-to-bottom, respectively, as a function of $t/t_0$. Colours and linestyles are the same as in \autoref{fig:Rm_energy_balance}. Smoothing splines are shown overlayed to the data to reveal the general trends in the evolution.}
            \label{fig:char_mag}
    \end{figure}  
    
\section{Probing the structure of the magnetic energy}\label{sec:magnetic_field_scales}
    In this section, we define three characteristic scales of the magnetic field that explore the correlation structure of the magnetic energy. The first scale of interest is the integral scale of the magnetic field, $\kcor$, which is computed directly from the 1D energy spectrum,
    \begin{align}\label{eq:correlation_scale}
        \kcor = \left(\int_0^{\infty} \d{k} \, k^{-1} \Exp{|\vecB{b}(\vecB{k})|^2}{\theta} \middle/ \int_0^{\infty}\d{k} \, \Exp{|\vecB{b}(\vecB{k})|^2}{\theta} \right)^{-1},
    \end{align}
    and probes the characteristic size of the largest field correlation. In the kinematic stage $\kcor \sim \Rm^{1/2} \sim \keta$ \citep{Galishnikova2022_saturation_and_tearing,Kriel2022_kinematic_dynamo_scales}. The next scale we define is the peak scale of the spectrum,
    \begin{align}\label{eq:peak_scale}
        \kp = \text{argmax}_{k}\left[ \mathcal{I}\left\{ \Exp{|\vecB{b}(\vecB{k})|^2}{\theta} \right\} \right],
    \end{align}
    where $\mathcal{I}\left\{\hdots\right\}$ is a cubic interpolation operator applied to the spectrum, allowing us to compute $\kp$ in a model-free and bin-free fashion. In the kinematic regime, $\kp \sim \keta$ \citep{Kriel2022_kinematic_dynamo_scales}. Also note that because $\kp\sim \keta$, $\kp\sim \kcor$ (see \aref{appendix:cor_and_peak_scale} for a more detailed comparison of $\kp$ and $\kcor$), this tells us that correlations are being destroyed on $k<\keta$, all through the Kazantsev interval ($k^{3/2}$, for which fields are folded; \citealt{Schekochihin2004_critical_dynamo}) of the spectrum, and piling up into folds around $\keta$. As the dynamo approaches the saturated stage, $\kp$ shifts back to lower $k$ modes, $\kp\ll \keta$ (possibly at the resistive dynamical timescale, \citealt{Schekochihin2002_saturation_evolution}) consistent with previous theory \citep{Schekochihin2002_saturation_evolution,Xu2016_dynamo,McKee2020} and simulations \citep{Seta2020_saturation_high_Pm,seta2021saturation}, and correlations are able to develop $k<\keta$. The third is the microscale of the magnetic field, $\ku$ (a scale analogous to the Taylor microscale or the rms scale of the field),
    \begin{align}\label{eq:micro_scale}
        \ku =& \left(\frac{\Exp{|\nabla\otimes\vecB{b}|^2}{\V}}{\Exp{\b^2}{\V}}\right)^{1/2},\\
            =& \left(\int_0^{\infty} \d{k} \, k^{2} \Exp{|\vecB{b}(\vecB{k})|^2}{\theta} \middle/ \int_0^{\infty} \d{k} \Exp{|\vecB{b}(\vecB{k})|^2}{\theta} \right)^{1/2},
    \end{align} 
    by Parsevel's theorem. This scale gives us a characteristic size of the magnetic field gradients, i.e., structures in the magnetic field that are not smoothed out by dissipation. In \citet{Kolmogorov1941} turbulence, the energy cascade is defined on the interval $\kcor > k > \ku$. 
        
    First, we annotate the correlation and microscale onto \autoref{fig:averged_ps}. The separation between the scales defines the range of $k$ within the turbulence cascade. This range is limited, only over a few $k$ modes, but this is the nature of doing turbulence studies in the presence of limited grid resolution, which in turn limit the values for $\Re$ and $\Rm$ and the separation between injection and dissipation in the plasma. As $\Rm$ increases, we see that $\keq$ (shown in \autoref{fig:k_eq}) moves closer to $\kcor$. We hypothesise that in the $\Rm\rightarrow\infty$ limit, $\keq \rightarrow \kcor$, in turn making $(\emag/\ekin)_{\text{sat}}$ independent of $\Pm$, which is the usual assumption for the models involving the integral energies \citep[e.g.,][]{Schober2015_saturation_of_turbulent_dynamo}. This is because, following the phenomenology presented in \citet{Galishnikova2022_saturation_and_tearing}, $\keq$ describes the largest scales that are being significantly fed by the kinetic energy modes, and naturally, they ought to support a correlated structure. Therefore, if $\keq < \kcor$ are significantly coupled to the kinetic energy modes, $\keq\sim\kcor$ as the turbulence correlates and grows them. 
        
    Now we explore these scales throughout the whole temporal evolution of the simulations. In \autoref{fig:char_mag} we show $\kcor$, $\ku$ and $\kp$ as a function of $t/t_0$, coloured in the same fashion as \autoref{fig:Rm_energy_balance}, and with different linestyles for different $\Rm$ experiments. For fixed $\Rm$, the value of the scales move to the same saturated state, as we showed previously in \autoref{fig:averged_ps}. However, similarly to the integral statistics, the journey towards saturation is very different for the different initial magnetic field strengths, similar to what we saw in the time-dependent energy spectra, \autoref{fig:time_evolution_spectra}. 
    Firstly, for the \texttt{weak} experiments (black) all three scales move from low-$k$ to high-$k$ modes in the kinematic regime, $0 \leq t/t_0\leq20$. This represents the whole spectra shifting to high-$k$, where $\kp \sim \kcor \sim k_{\eta}$ \citep{Schekochihin2004_dynamo,Xu2016_dynamo,Kriel2022_kinematic_dynamo_scales,Galishnikova2022_saturation_and_tearing}. \citet{Xu2016_dynamo} predicts that $\kp$ starts to move to lower $k$-modes as the dynamo approaches the saturation, which we see happen for $t/t_0\gtrsim20$. The extent of the change between the scales in the kinematic and saturated regime is larger with increasing $\Rm$.  
        
    In contrast to the \texttt{weak} simulations, the \texttt{strong} simulations (aqua) are dominated by low-$k$ modes; ($\kcor \sim \kp \sim 1/L$), hence have minimal field line curvature, and too with only large-scale magnetic field gradients $(\ku \sim 2/L)$, and therefore large-scale Lorentz force and dissipation (e.g., both \autoref{eq:momentum}, \autoref{eq:induction} strictly rely upon gradients) but after a critical $t/t_0$, which we will discuss in Paper II, the magnetic field structure hastily responds to the decaying field, before finding the saturation beyond $t/t_0 \gtrsim 200$. We can see from \autoref{fig:time_evolution_spectra}, that this is roughly at the $t/t_0$ where the high-$k$ modes are growing through the dynamo action (operating on dynamical timescales on those scales; \citealt{Haugen2004_dynamo_mach_and_crit_rm}), shortly before the low-$k$ modes have all but decayed. Naturally, once $\ku$ moves to the high-$k$ modes, the regular turbulent cascade in the magnetic field (see Figure~8, row BB in \citealt{Grete2017_shell_models_for_CMHD}) can proceed, and the Ohmic diffusion that we set on the small scales can destroy the magnetic field. The key conclusion is that the overall magnetic field structure, including the correlation, magnetic peak energy and micro-scale all find the same values for a given $\Pm$, even though the journey there is completely different between the different $E_{\text{mag},0}/\ekin$ experiments.
        
\section{Implications and limitations}\label{sec:implications_and_limitations}

    \subsection{Implications}
        For $\emagb \ll \ekinb$, the turbulent dynamo will grow a magnetic field exponentially fast into a saturated state, for $\emagb \sim \ekinb$ the dynamo will maintain the magnetic field in a saturated state, and for $\emagb \gg \ekinb$, $\emagb$ field will decay into a saturated state. The saturation in all of these processes is exactly the same, and hence the physics of the saturation does not depend upon any initial structure and amplitudes generated by previous stages, kinematic or otherwise, that happen before the saturation, e.g., the $\propto k^{3/2}$ spectrum in the kinematic stage or by stretching/twisting/folding \citep[e.g.,][]{Kazantsev1968,Schekochihin2004_dynamo,Galishnikova2022_saturation_and_tearing,Seta2020_saturation_high_Pm,2023Sur_b_field_pressure_saturation,Kempski2023_b_field_reversals}. Moreover, it means that the saturation in isotropic MHD turbulence can be studied with any set of initial conditions, not necessarily needing to go through the other two growth stages. Of course, this does not make the stretch/twist/fold processes in those growth stages any less important for unravelling the details of magnetic field growth and maintenance. Moreover, for studies focused on the saturated state, it might be advantageous to study the plasma with $\emagb \lesssim \ekinb$ initial conditions. On the other hand, for $\emagb \gg \ekinb$, i.e., capturing the magnetic field terms in an amplified state, this provides a different perspective on the journey towards the saturated stage, which may lead to some insight into the magnetic processes that then have to balance with the turbulence to create the final steady state.
        
        In the context of astrophysics, we have studied the saturation of isotropic blobs of gas, which could be any blob of plasma where the size-scale is significantly larger than the magnetic field correlation length \citep{Beattie2022_ion_alfven_fluctuations,Beattie2022energy_balance}. Because the $\emagb \ll \ekinb$ dynamo grows fast, $\Exp{\b^2}{\V} \propto \exp\left\{ \gamma t \right\}$, and the same saturation can be reached by any $\emagb/\ekinb$ configuration, this makes the $\emagb/\ekinb$ set by the turbulent dynamo a sensible lower bound for estimating the magnetisation in turbulent plasmas across the modern Universe. Hence, if one can measure the level of turbulence, and estimate the $\Pm$ of plasma, one should in principle be able to invoke dynamo theory to get both a steady state and lower bound of the magnetisation.

    \subsection{Limitations}
        We probe only a limited set of parameters for $\Rm$ (and $\Re$), which will certainly dictate both how fast the dynamo grows into the saturated state and importantly, how fast or slow the magnetic field decays into the saturated state. For high $\Rm$, and high Lundquist number, $S = \Exp{v_A^2}{\V}^{1/2}L/\eta$, super-Alfv\'enic plasmoid instabilities in current sheets may cause fast reconnection and dominate the decay process \citep{Biskamp1986_magnetic_reconnection,Bhattacharjee2009_fast_reconnection,Uzdensky2010_plasmoid_reconnection,Hosking2021_reconnection_controlled_decay,Galishnikova2022_saturation_and_tearing,Dong2022_reconnection_mediated_cascade,Fielding2022_ISM_plasmoids}. Hence at higher $\Rm$ we may find that the exponential decay functions that we observed in \autoref{fig:Rm_energy_balance} turn into power-law decay might form, as shown in decaying MHD experiments. However, this kind of decay is significantly different from a regular decay experiment, since the turbulence is continuously driven and is able to grow low-$k$ modes, and the Lundquist number is sufficiently high ($\gtrsim 10^4$ at the start of the decaying sims, due to the very strong magnetic field) but the Alfv\'en Mach number is sufficiently low. Therefore this is an interesting growth and decay regime to further explore, which we do in Paper II.
        
        Naturally, the parameter space for MHD turbulence is large, and the same goes for the dynamo (e.g., helical, $\omega$, $\alpha$, etc., and combinations thereof; \citealt{Brandenburg2005_dynamo_review,Rincon2019_dynamo_theories}). A plethora of dynamos exist, and we have focused solely on the turbulent dynamo in isotropic non-helical MHD turbulence in a triply periodic box. Furthermore, we do not explore the case where there is a mean magnetic field (on the large scales), which fundamentally changes the saturation, suppressing high-$k$ mode growth \citep{Federrath2016_dynamo,Beattie2022energy_balance,Skalidis2023_sub_alf_dynamics}.
              
\section{Summary and conclusions}\label{sec:conclusion}
    Using an ensemble of isotropic, visco-resistive, three-dimensional, non-helical, magnetohydrodynamic turbulence simulations we study the statistical properties of the small-scale turbulent dynamo saturation at different magnetic Prandtl number $\Pm$ and initial magnetic field amplitude and structure. For a given $\Pm$, in the saturated state of the turbulent dynamo, we find that the integral energies, energy spectra, and characteristic scales of the magnetic field are attracted to the same values, regardless of the initial magnetic field configuration (structure or amplitude), and even if initially $\emagb \gg \ekinb$ and the magnetic field is forced to decay into the saturation. This suggests that for a specific $\Pm$, the structure and energy in the magnetic field are controlled solely by the turbulence. Because (1) the long-term behaviour of the turbulent dynamo is invariant to the history of the magnetic field and (2) the kinematic stage of the dynamo is fast, we highlight how the $\emagb/\ekinb$ saturation value could be interpreted as an estimate for both the steady state (regardless of initial seed field) and a reasonable lower bound of the magnetisation in a turbulent plasma.

\section*{Acknowledgements}
    We thank the anonymous referee who helped increase the clarity and strength of the arguments presented in this study. We thank Siyao Xu, Jim Stone, Bart Ripperda, Archie Bott, Justin Kin Jun Hew, Brant Robertson, and Lachlan Lancaster for the general discussions regarding this work. 
    
    J.~R.~B.~acknowledges financial support from the Australian National University, via the Deakin PhD and Dean's Higher Degree Research (theoretical physics) Scholarships and the Australian Government via the Australian Government Research Training Program Fee-Offset Scholarship and the Australian Capital Territory Government funded Fulbright scholarship. C.~F.~acknowledges funding provided by the Australian Research Council (Future Fellowship FT180100495 and Discovery Project DP230102280), and the Australia-Germany Joint Research Cooperation Scheme (UA-DAAD). We further acknowledge high-performance computing resources provided by the Leibniz Rechenzentrum and the Gauss Centre for Supercomputing (grants~pr32lo, pn73fi, and GCS Large-scale project~22542), and the Australian National Computational Infrastructure (grant~ek9) and the Pawsey Supercomputing Centre (project~pawsey0810) in the framework of the National Computational Merit Allocation Scheme and the ANU Merit Allocation Scheme. P.~M.~acknowledges this work was in part performed under the auspices of the U.S. Department of Energy by Lawrence Livermore National Laboratory under contract DE-AC52-07NA27344, Lawrence Livermore National Security, LLC.\\
    
    The simulation software, \textsc{flash}, was in part developed by the Flash Centre for Computational Science at the Department of Physics and Astronomy of the University of Rochester. \textsc{TurbGen} \citep{Federrath2010,Federrath2022_turbulence_driving_module} for the turbulent forcing function, $\vecB{f}$. Data analysis and visualisation software used in this study: \textsc{C++} \citep{Stroustrup2013}, \textsc{numpy} \citep{Oliphant2006,numpy2020}, \textsc{matplotlib} \citep{Hunter2007}, \textsc{cython} \citep{Behnel2011}, \textsc{visit} \citep{Childs2012}, \textsc{scipy} \citep{Virtanen2020},
    \textsc{scikit-image} \citep{vanderWalts2014}.

\section*{Data Availability}
    The data underlying this article will be shared on reasonable request to the corresponding author.

\bibliographystyle{mnras.bst}
\bibliography{Jan2022.bib} 

\appendix
\section{Convergence Test}\label{appendix:convergence}
\begin{figure}
        \centering
        \includegraphics[width=\linewidth]{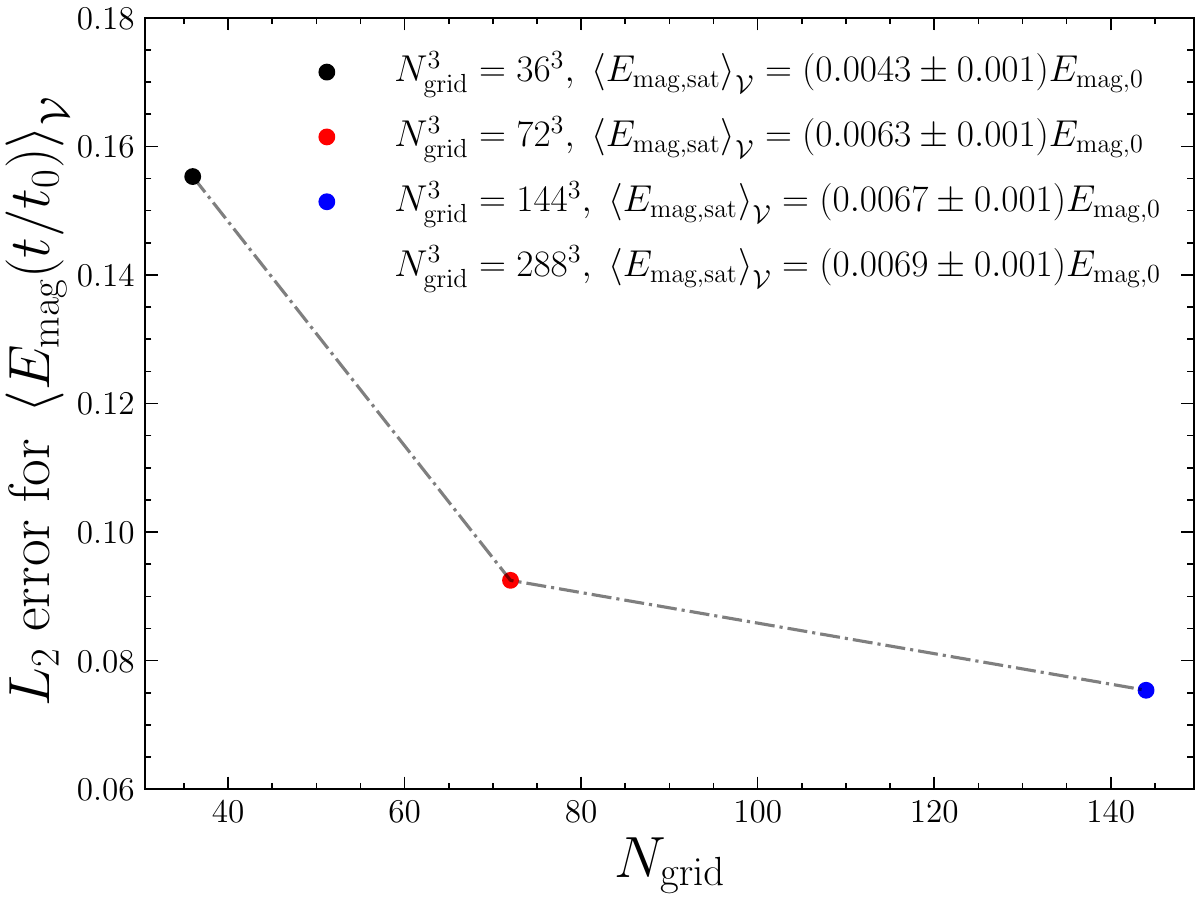}
        \caption{The $L_2$ error for $\emagb$ in the \texttt{strongPm2} experiments in the saturation, (see \autoref{tab:sims}) at grid resolutions $N_{\text{grid}}^3 = \left\{36^3, 72^2, 144^3\right\}$. }
        \label{fig:convergence}
    \end{figure}

    For a set of \texttt{strong}, $\Pm =2$ simulations listed in \autoref{tab:sims}, we perform the $L_2$ error between $\left\langle E_{\text{mag},k}(t/t_0) \right\rangle$ for $k \in \left\{36^3, 72^3, 144^3\right\}$ grid resolutions and $E_{\text{mag},288}(t/t_0)$,
    \begin{align}
        L_2\;\text{error} &= \| \left\langle E_{\text{mag},k}(t/t_0) \right\rangle - \left\langle E_{\text{mag},288}(t/t_0) \right\rangle \|_2, \\
        &= \left(\frac{1}{N}\sum^N_{\forall t/t_0} \bigg[\left\langle E_{\text{mag},k}(t/t_0) \right\rangle - \left\langle E_{\text{mag},288}(t/t_0) \right\rangle\bigg]^2\right)^{1/2},
    \end{align}
    computed in the saturated stage ($t/t_0 \geq 200$). We show the plot of the $L_2$ error as a function of linear grid resolution in \autoref{fig:convergence}, showing that the $L_2$ is a monotonically decreasing function, converging slowly towards the $288^3$ data. Likewise, we show the averaged saturation of the magnetic field energy $\left\langle E_{\text{mag,sat}} \right\rangle_{\V}$ in the legend, and $(\emagb/\ekinb)_{\text{sat}}$ in \autoref{tab:sims} for each of the resolutions, highlighting that by $144^3$, our results are converged within 1$\sigma$ for both quantities.

\section{Energy equipartition scale calculation}\label{appendix:k_eq}
    We define the energy equipartition mode $\keq(t/t_0)$ in \autoref{fig:k_eq} as
    \begin{align}\label{eq:k_eq_define}
        \left\{k(t/t_0) | \mathcal{I}\left\{\emag(k, t/t_0)\right\} - \mathcal{I}\left\{\ekin(k, t/t_0)\right\} = 0 \ \right\},
    \end{align}
    where $\mathcal{I}\left\{\emag(k, t/t_0)\right\}$ and $\mathcal{I}\left\{\ekin(k, t/t_0)\right\}$ are the interpolated energy spectra. For each $t/t_0$ we have a spectrum of ordered $N$ $k(t/t_0)$ that satisfy \autoref{eq:k_eq_define},
    \begin{align}
        \left\{ k_{\text{eq},i} \right\} = \left\{ k_{\text{eq},1}, k_{\text{eq},2}, k_{\text{eq},3}, \hdots k_{\text{eq},N} \right\},
    \end{align}
    due to fluctuations in modes deep in the numerical dissipation regime (see bottom panel of \autoref{fig:time_evolution_spectra}). Hence, to be consistent with the relevant equipartition scale in e.g., \citep{Galishnikova2022_saturation_and_tearing}, we take the root at the lowest-$k$ modes,
    \begin{align}\label{eq:k_eq}
        \keq \equiv \min\left\{ k_{\text{eq},i} \right\},
    \end{align}    
    for each $t/t_0$. Naturally, this provides us with a length scale where the plasma transitions from being dominated by $\ekin(k)$ to being dominated by $\emag(k)$ modes in the plasma, as we show directly in \autoref{fig:averged_ps}. $k_{\rm eq}$ appears later in the time-evolution of the simulation, and based on \autoref{fig:k_eq} one can see that this changes for different $\Rm$, so we begin plotting the scale upon the first appearance of it in the simulations.
        
    \begin{figure*}
        \centering
        \includegraphics[width=\linewidth]{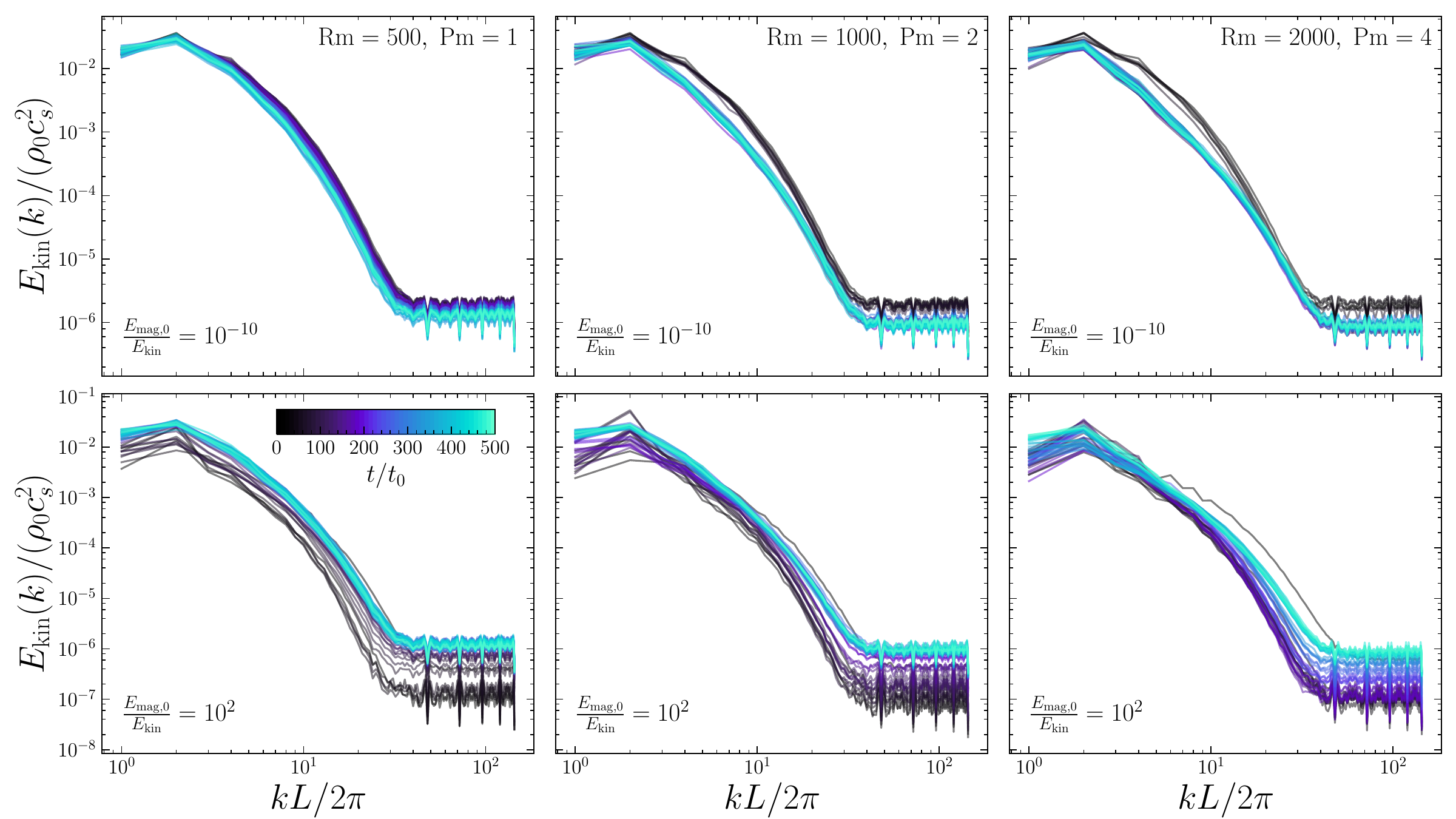}
        \caption{The same as \autoref{fig:time_evolution_spectra} but for the kinetic energy spectra. \textbf{Top:} showing that as the kinetic energy comes close to saturation the spectra reduce in energy around the modes close to the equipartition scale for the \texttt{weak} experiments (top row). \textbf{Bottom:} the kinetic energy is suppressed (work is done by the sub-Alfv\'enic magnetic field) on all scales during the magnetic decay in the \texttt{strong} experiments.}
        \label{fig:kinetic_spectra}
    \end{figure*}
    
    \begin{figure*}
        \centering
        \includegraphics[width=\linewidth]{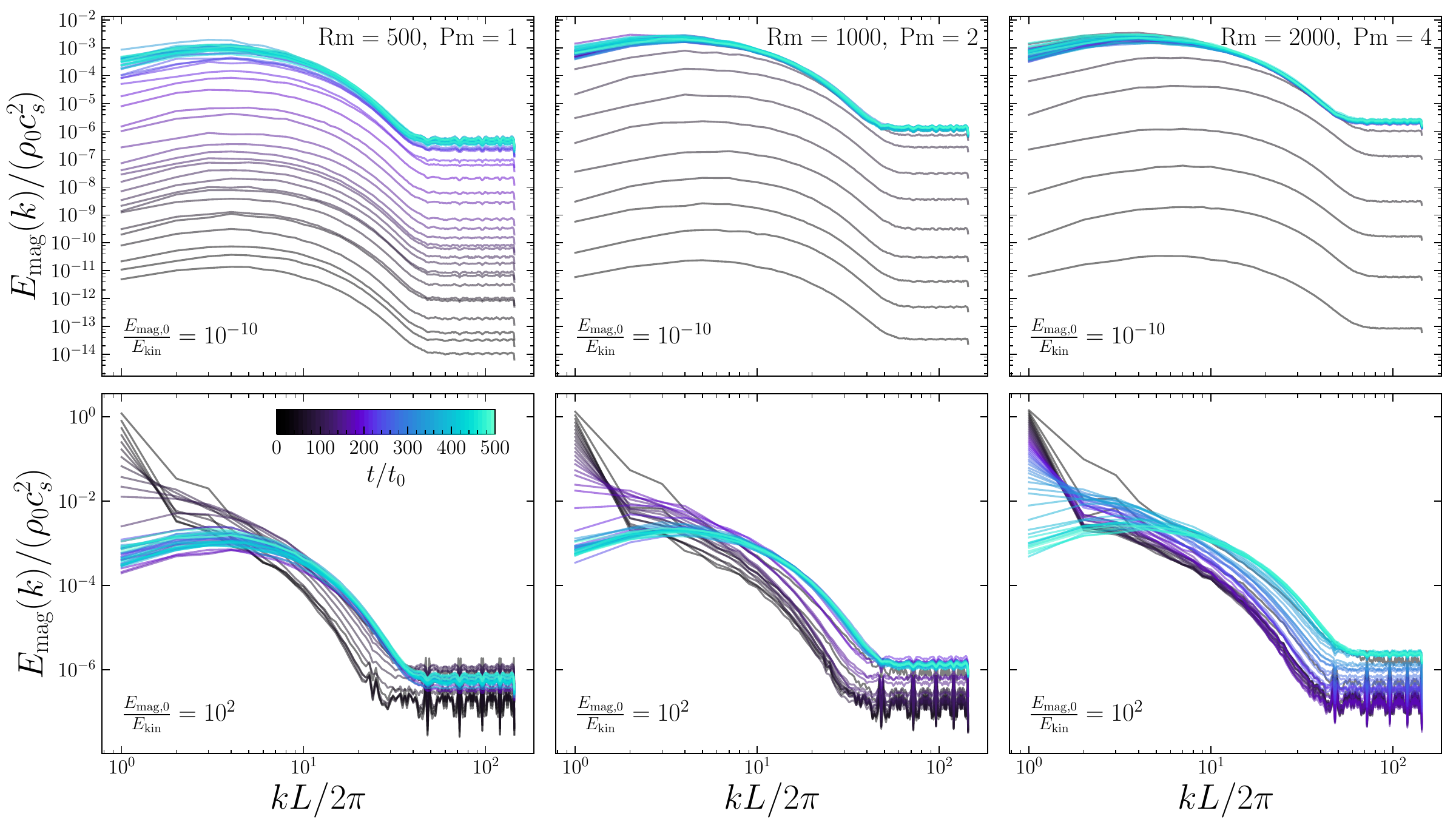}
        \caption{The same as \autoref{fig:kinetic_spectra} but for the magnetic energy spectra.}
        \label{fig:magnetic_spectra}
    \end{figure*}

\section{Energy spectra}\label{appendix:kinetic_spectra}
    In \autoref{sec:spectrum} we explored the time-evolution of the $\emag(k)/\ekin(k)$ ratio and the saturated $\emag(k)$ and $\ekin(k)$ spectra, however, neither of these plots clearly showed details of the time-evolution of $\ekin(k)$ or $\emag(k)$ separately. In \autoref{fig:kinetic_spectra} we show $\ekin(k,t)$, using the same panel configuration as in \autoref{fig:time_evolution_spectra}, and likewise for $\emag(k,t)$ in \autoref{fig:magnetic_spectra}. 
    
    Notably, for the \texttt{strong} experiments (bottom panel) in \autoref{fig:kinetic_spectra} $\ekin(k,t)$ slowly grows on all $k$-modes, until it finally reaches a saturated state (corresponding to the volume-weighted $\M = 0.5$). Compare this with the top-panel of \autoref{fig:kinetic_spectra} and previous dynamo studies, where $\ekin(k,t)$ reaches saturation within a few turnover times \citep[e.g.,][]{Kriel2022_kinematic_dynamo_scales}.   indicating that as $\emag$ decays, there is a conversion into $\ekin$ (possibly via very slow reconnection events or the Lorentz force slowly smoothing out the gradients in the strong magnetic field). The classical dynamo experiments (top panel) show a shallowing kinetic energy spectrum as the transition from the kinematic (black) regime, to the saturated (aqua) regime, opposite to what is expected to happen based on the scale-by-scale equipartition between the magnetic tension and strain \citep{Galishnikova2022_saturation_and_tearing}. Now we turn our attention to \autoref{fig:magnetic_spectra}. As we discuss throughout the main text, $\emag(k,t)$ is initially dominated by low-$k$ modes, which decay as the high-$k$ modes that are coupled to the turbulence grow. This facilitates simultaneous growth and decay in different parts of the energy spectra. 

\section{Magnetic spectra scale correlations}\label{appendix:cor_and_peak_scale}
    \begin{figure}
        \centering
        \includegraphics[width=\linewidth]{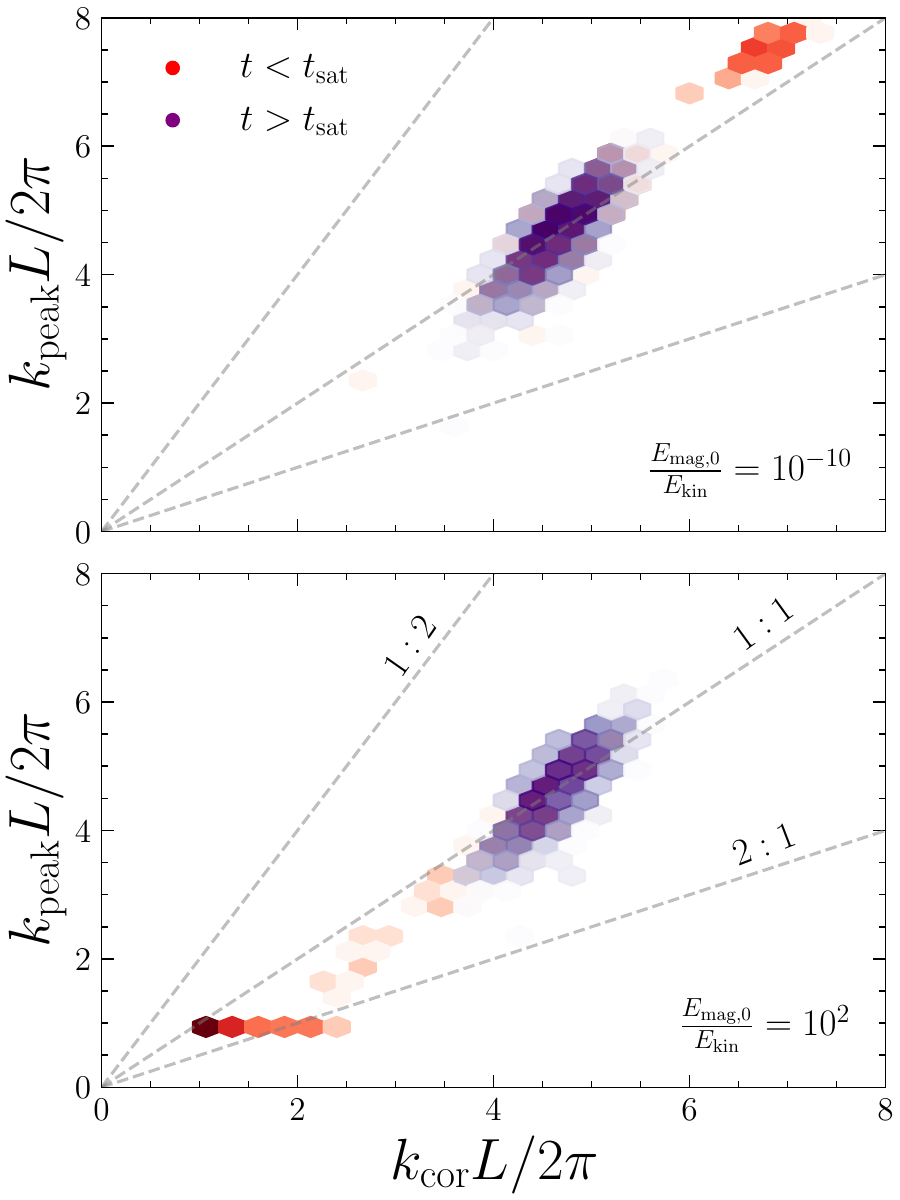}
        \caption{The 2D distribution of $\kp$ as a function of $\kcor$ for the $\Pm=4$ \texttt{weak} (top) and \texttt{strong} (bottom) simulations, coloured by red for $t/t_0$ that are less than $t_{\rm sat}/t_0$, where $t_{\rm sat}$ is the time it takes to get into the saturated state of the small-scale dynamo and purple for $t/t_0$ in the saturated state. The transparency for each bin is proportional to the $t/t_0$ spent at that $\kp$/$\kcor$ value. Both simulations exhibit very close to $1:1$ scaling between the scales, regardless of the dynamo stage.}
        \label{fig:k_cor_k_peak_2d}
    \end{figure}
    
    \begin{figure}
        \centering
        \includegraphics[width=\linewidth]{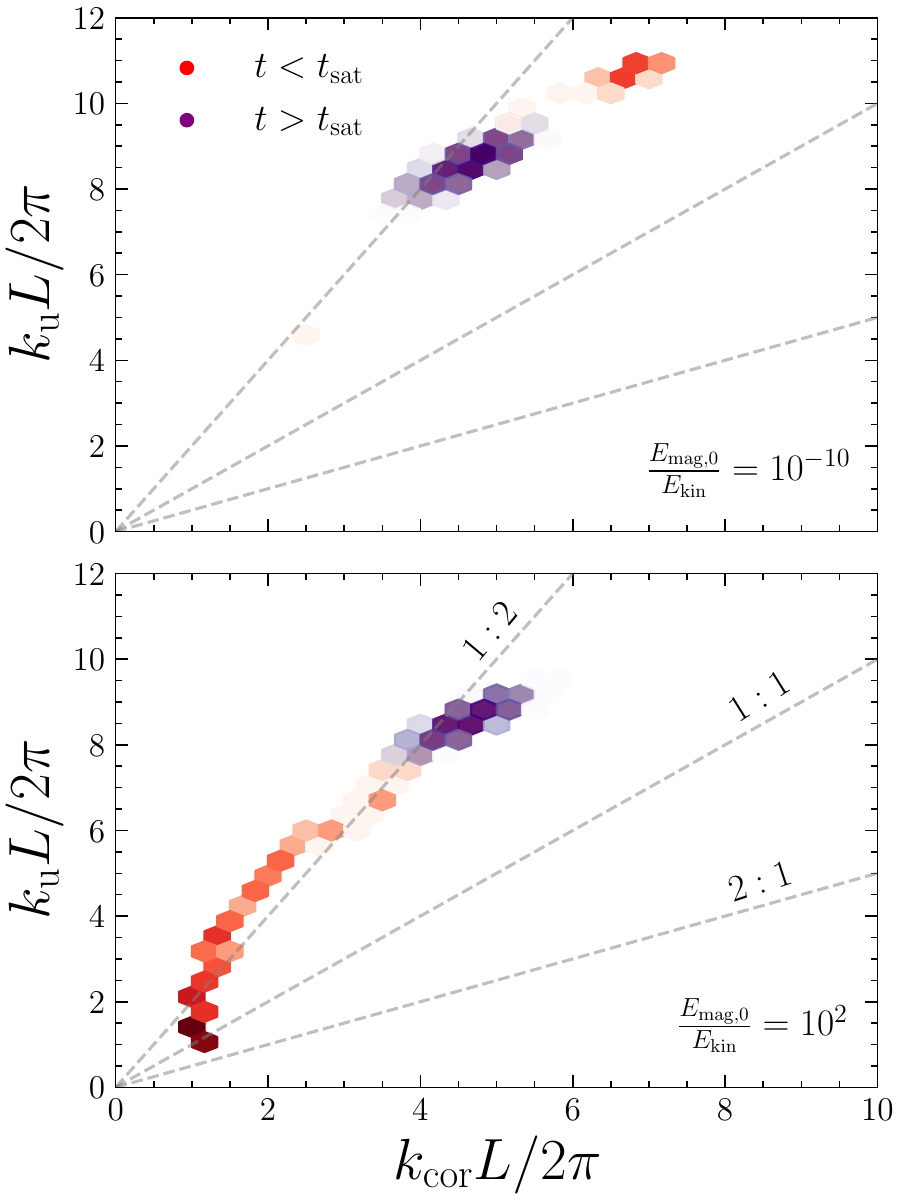}
        \caption{The same as \autoref{fig:k_cor_k_peak_2d}, but for $\ku$ (\autoref{eq:micro_scale}) as a function of $\kcor$.}
        \label{fig:k_cor_k_u_2d}
    \end{figure}
    \citet{Schekochihin2004_dynamo} and \citet{Galishnikova2022_saturation_and_tearing} assume that $\kcor$ (\autoref{eq:correlation_scale}), the correlation of scale of $\emag(k)$ is proportional to peak energy scale, $\kp$ (\autoref{eq:peak_scale}), in the kinematic stage of the SSD. Qualitatively, \autoref{fig:char_mag} shows that this seems true in not only the kinematic stage, but the nonlinear and saturated stage of the dynamo, and also in the decaying stage of the \texttt{strong} experiments. Here we show two representative plots in \autoref{fig:k_cor_k_peak_2d} to quantitatively confirm that this is indeed the case, no matter what the state of the magnetic field. 
    
    In the top panel of \autoref{fig:k_cor_k_peak_2d} we show the 2D histogram of $\kp$-$\kcor$ for the whole $t/t_0$ in the $\Pm=4$ \texttt{weak} simulation and in bottom we show the corresponding \texttt{strong} simulation. The colouring indicates whether the data for the scales is before the saturation (red) or after (purple). As we showed in \autoref{fig:char_mag}, both of the scales move to lower-$k$ in the saturated state compared to the kinematic stage in the \texttt{weak} calculations, and the opposite for the \texttt{strong} calculations. The opacity in each hexagonal bin corresponds to the amount of data in that bin, which in turn corresponds to the amount of time spent at that $(\kcor,\kp)$ value. This shows that most of the time, whether the magnetic field is growing, decaying, or in the saturated state, $\kp = \kcor$. During the evolution between the stages this never deviates by a factor of $\gtrsim 2$, as we show with the $1:2$ and $2:1$ lines. 
    
    We make the same plot for $\ku$-$\kcor$ in \autoref{fig:k_cor_k_u_2d}. It shows that for both the \texttt{weak} and \texttt{strong} experiments, $\ku \approx 2\kcor$ through the kinematic phase (top panel) and decay phase (bottom panel), $t < t_{\rm sat}$. As $t \rightarrow t_{\rm sat}$, $\ku$ and $\kcor$ move to smaller wavenumbers for the \texttt{weak} experiments, and higher for the \texttt{strong} experiments, all whilst maintaining a similar $\ku \approx 2\kcor$ relation. We suspect that this separation increases with $\Rm$, however, the main point is that there is separation between the $\kcor$ and $\ku$ that is maintained throughout the entire growing dynamo and decaying process. 

\bsp	
\label{lastpage}
\end{document}